\documentclass[preprint,superscriptaddress]{revtex4}

\usepackage{amsmath}
\usepackage{epsfig}
\usepackage{graphics}
\usepackage{graphicx}
\usepackage[colorlinks=true, citecolor=blue, urlcolor = blue, linkcolor= red, bookmarks=true]{hyperref}

\begin{document}

\title{Shadow images of Kerr-like wormholes}

\author{Muhammed Amir}
\email{amirctp12@gmail.com}
\affiliation{Astrophysics and Cosmology Research Unit,
 School of Mathematics, Statistics and Computer Science,
 University of KwaZulu-Natal, Private Bag X54001,
 Durban 4000, South Africa}

\author{Kimet Jusufi}
\email{kimet.jusufi@unite.edu.mk}
\affiliation{Physics Department, State University of Tetovo, 
  Ilinden Street nn, 1200, Tetovo, North Macedonia}
\affiliation{Institute of Physics, 
  Faculty of Natural Sciences and Mathematics,
  Ss. Cyril and Methodius University,
  Arhimedova 3, 1000 Skopje, North Macedonia}
 
\author{Ayan Banerjee}
\email{ayanbanerjeemath@gmail.com}
\affiliation{Astrophysics and Cosmology Research Unit,
 School of Mathematics, Statistics and Computer Science,
 University of KwaZulu-Natal, Private Bag X54001,
 Durban 4000, South Africa}
 
\author{Sudan Hansraj}
\email{hansrajs@ukzn.ac.za}
\affiliation{Astrophysics and Cosmology Research Unit,
 School of Mathematics, Statistics and Computer Science,
 University of KwaZulu-Natal, Private Bag X54001,
 Durban 4000, South Africa}

\date{\today }

\begin{abstract}
Investigations  of shadows of astrophysical entities constitute  a major source of insight into the 
evolution of compact objects. Such effects depend on the nature of the compact object and arise on 
account of the strong gravitational lensing  that casts a shadow on the bright background. We consider 
the Kerr-like wormhole spacetime (Phys.\ Rev.\ D 97:024040, 2018), which is a modification of the Kerr 
black hole that degenerates into wormholes for nonzero values of the deviation parameter $\lambda^2$. 
The results suggest that the Kerr spacetime can reproduce far away from the throat of the wormhole. 
We obtain the shapes of the shadow for the Kerr-like wormholes and discuss the effect of the spin $a$, 
the inclination angle $\theta_0$, and the deviation parameter $\lambda^2$ on the size and nature of the 
shadow. As a consequence, it is discovered that the shadow is distorted due to the spin as well as 
the deviation parameter and the radius of the shadow decreases with $\lambda^2$ if the ADM mass of the Kerr-like wormholes is considered.
\end{abstract}

\maketitle

\section{Introduction}
\label{sec:intro}
In the context of general relativity wormholes act as tunnel-like structures  connecting two different 
regions in spacetime. This phenomenon necessitates solving Einstein's equations in a reversed manner 
to obtain solutions admitting wormholes. Studies on this topic can be divided into two classes: one 
relating to the Euclidean wormholes \cite{Ruz:2013hfa,ArkaniHamed:2007js,Hawking:1987mz,Hawking:1988ae} 
and the other concerned with the Lorentzian ones. Historically, the notion of wormhole physics may be 
traced back to 1916 by Flamm \cite{Flamm}, soon after the discovery of the Schwarzschild solution. 
Einstein and Rosen \cite{Einstein:1935tc} proposed a mathematical construction in order to eliminate 
coordinate or curvature singularities, eventually introduced a bridge-like structure connecting two 
identical sheets known as the \emph{Einstein-Rosen bridges}. This mathematical  representation of 
physical space was not successful as a model for particles rather it emerged as the prototype wormhole 
in gravitational physics. Subsequently, Wheeler \cite{WH,Fuller:1962zza} revived the subject in the 
1950s where the Kruskal-Szekeres coordinates were employed to describe the geometry of the 
Schwarzschild wormhole, although these wormholes were at the quantum scale. Wheeler coined the 
term \emph{wormhole} and later his solutions were transformed into Euclidean wormholes by Hawking 
\cite{Hawking:1988ae} and others.

Interest in traversable Lorentzian wormholes has been on the incline recently  following  the 
stimulating work of Morris and Thorne \cite{MM}. In this proposal it is possible for  observers to  
freely traverse wormholes and  time travel is admissable. They initially introduced concepts as 
pedagogic tools for teaching general relativity where the Einstein field equations are solved by 
constructing the spacetime metric first and then deducing the stress-energy tensor components. In order 
to construct such traversable wormhole solutions demands the existence of exotic matter whose 
stress-energy tensor components violate the null, weak, and the strong energy conditions 
\cite{MM,Morris:1988tu}, at least in a neighborhood of the wormhole throat. For a more detailed review 
the reader may consult  \cite{MV,Lobo:2007zb} and references therein. While the introduction of exotic 
matter may appear unpalatable such matter fields exist in the quantum theory of the gravitational field 
(namely the Casimir effect and Hawking evaporation \cite{Klinkhammer}) as well as in scalar-tensor 
theories. In fact wormhole geometries violate the averaged energy conditions \cite{MV}.

Historically various attempts have been made to avoid appealing to exotic matter but all were in vain 
within the context of general relativity. An advance in this discussion was  to consider minimizing the 
violation of the energy conditions, namely, the \emph{volume integral quantifier} 
\cite{Visser:2003yf,Kar:2004hc},
which quantifies the total amount of  matter violating the energy conditions. In this regard  Nandi 
\emph{et al.} \cite{Nandi:2004ku} considered an exact integral quantifier for matter violating the 
averaged null energy condition (ANEC). Kuhfittig \cite{Kuhfittig:2003pu} reported  that the region 
consisting of exotic matter can be made arbitrarily small. Therefore, it is not an easy task to search 
for promising candidates for exotic matter because all classical forms of matter are believed to obey 
the standard energy-conditions. Dark energy models with exotic equations of state have been proposed 
due to the formation and the evolution of the present structures in the Universe. While observational and 
experimental support is presently lacking, strong theoretical arguments for dark matter have been advanced 
principally to address the anomalous accelerated expansion of the universe problem. Consequently, much 
effort has been made to obtain stable wormhole solutions within these new paradigms. These include models 
supported by the phantom energy \cite{Lobo:2005us,Lobo:2005yv,Zaslavskii:2005fs}, the cosmological constant 
\cite{Rahaman:2006xa,Lemos:2003jb,Cataldo:2008ku}, generalized or modified Chaplygin gas 
\cite{Chakraborty:2007zi,Jamil:2008wu,Kuhfittig:2009mx}. However, in some alternative or modified theories 
of gravity, the requirement of exotic matter can be minimized or even completely avoided. For instance, it 
was shown that the matter threading the wormhole throat satisfies all of the energy conditions by Harko 
\emph{et al.} \cite{Harko:2013yb}, and it is the higher order curvature terms, interpreted as a 
gravitational fluid, that support the wormhole geometry. In \cite{MontelongoGarcia:2010xd}, wormhole 
solutions have also been found with matter satisfying the null energy  condition were nonminimal 
curvature-matter coupling in a generalized $f(R)$ modified theory of gravity. These kind of solutions were 
also found in various modified theories, e.g., Einstein-Gauss-Bonnet \cite{Mehdizadeh:2015jra}, 
Born-Infeld gravity \cite{Shaikh:2018yku}, Einstein-Cartan \cite{Mehdizadeh:2017tcf}, and trace of the 
energy-momentum tensor squared gravity \cite{Moraes:2017dbs}.

The detection of gravitational waves (GWs) by the advanced Laser Interferometer Gravitational Wave 
Observatory (aLIGO) \cite{TheLIGOScientific:2014jea} and Advanced Virgo \cite{TheVirgo:2014hva} detectors 
offers exciting opportunities towards the study of the nature of black holes. The detection of GWs from 
five binary black holes (BBHs) demonstrated that stellar-mass black holes really exist in our universe. 
However, the complete structure of the spacetime inside the light ring and near-horizon is still opaque 
from the current observations. Thus, the physics in strong gravitational field, e.g., near black holes is 
still an important topic not only in theoretical but also observational investigations.
Analysing the physical nature of black holes requires the actual detection of the event horizon. To confirm 
the presence of event horizons a number  of tests have been performed that help reveal exciting physics and 
astrophysics possibilities \cite{Broderick:2009ph,Broderick:2015tda,Narayan:2008bv}. The evidence is 
compelling however not conclusive \cite{Abramowicz:2002vt}. Another important phenomenological feature of 
black holes is their shadow. The concept of the rotating black hole shadow was suggested by Bardeen 
\cite{Bardeen:1973gb}, with the idea of the optical image of black hole appearing due to the strong 
gravitational lensing effect. Black holes are  expected to cast shadows on the bright background. Current 
observations suggest that most galaxies contains supermassive black holes at their centers 
\cite{Rees:1984si,Kormendy:1995} and that galaxies are rotating. This means that the black hole at the 
center of a galaxy also possesses a spin and in this environment it is interesting to investigate the 
nature of the black hole by analyzing the shadow that they cast on the bright background. Motivated by 
present observational missions \cite{Doeleman}, the study of the black hole shadow has received  
significant attention. Various configurations of black hole solutions, such as the shadow of the 
Schwarzschild black hole \cite{Synge:1966,Luminet:1979}, have been investigated. The investigation of the 
shadow for several black holes in the context of modified theories of gravity and higher-dimensional 
theories have received attention in 
\cite{Takahashi:2005hy,Bambi:2008jg,Hioki:2009na,Wei:2013kza,Bambi:2010hf,Amarilla:2010zq,Amarilla:2011fx,Amarilla:2013sj,Yumoto:2012kz,Abdujabbarov:2012bn,Atamurotov:2013sca,Li:2013jra,Cunha:2015yba,Johannsen:2015qca,Abdujabbarov:2016hnw,Amir:2016cen,Kumar:2017vuh,Singh:2017xle,Papnoi:2014aaa,Amir:2017slq,Singh:2017vfr}. 
Shadow casting by the Kerr-like black holes in modified gravity theory has been addressed recently in 
\cite{Wang:2018prk}. Different approaches to investigate the black hole shadow has been addressed so far in 
the literature, e.g., photon regions based description of shadow \cite{Grenzebach:2014fha}, Hioki and Maeda 
approach to calculate the observables of shadow \cite{Hioki:2009na}, new method to perform 
general-relativistic ray-tracing for shadow images \cite{Younsi:2016azx}, coordinate-independent method 
\cite{Abdujabbarov:2015xqa}, in presence of plasma \cite{Perlick:2015vta}, analytic description of shadow 
\cite{Tsupko:2018apb}.

The main motivation of the paper is to investigate the shadows of Kerr-like wormholes obtained by Bueno 
\emph{et al.} \cite{Bueno:2017hyj}, where the Kerr black hole turns out to be a wormhole solution that 
reproduces Kerr's spacetime away from the throat. The same procedure outlined was followed by  Damour and 
Solodukhin to obtain the Schwarzschild-like wormhole \cite{Damour:2007ap}. The detection of ringdown 
frequencies from the black holes provides precise tests that astrophysical black holes are verily described 
by the Kerr spacetime. Therefore, we conduct an  analysis of shadow for the Kerr-like wormholes. In 
addition, it is expected that the shadow of the Kerr-like wormholes can be used to probe the true nature 
of wormholes by Very Long Baseline Interferometry (VLBI) observations. In this direction a recent paper 
has been devoted to the study of the shadow boundary associated with the outer spherical orbits 
\cite{Nedkova:2013msa} and later this work was  developed by \cite{Shaikh:2018kfv}. Ellis wormholes have 
been analyzed  by using the images of wormholes surrounded by optically thin dust in \cite{Ohgami:2015nra}. 
The shadow of rotating wormholes in a plasma environment has been studied in the literature by the authors 
\cite{Abdujabbarov:2016efm}. Recently, the shadow of a class of charged wormholes solutions in EMDA theory 
has been studied in \cite{Amir:2018szm}.

The paper is organized as follows: after a brief introduction in Sect.~\ref{sec:intro}, we review the 
Kerr-like wormhole solution and construct embedding diagrams to represent a wormhole in 
Sect.~\ref{spacetime}. In Sect.~\ref{geodesics}, we evaluate the geodesic equations and derive the photon 
trajectories around the wormhole in Sect.~\ref {orbit}. Section~\ref{shadow} is devoted  to constructing 
the shadow images of Kerr-like wormholes. A discussion on the results is included in Sect.~\ref{conclusion}.

\section{Kerr-Like wormholes and embedding diagrams}
\label{spacetime}
\begin{figure*}
    \includegraphics[scale=0.98]{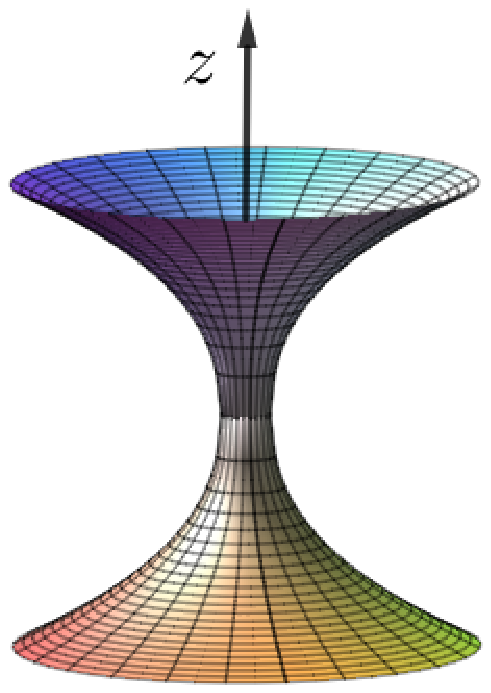}
    \includegraphics[scale=0.91]{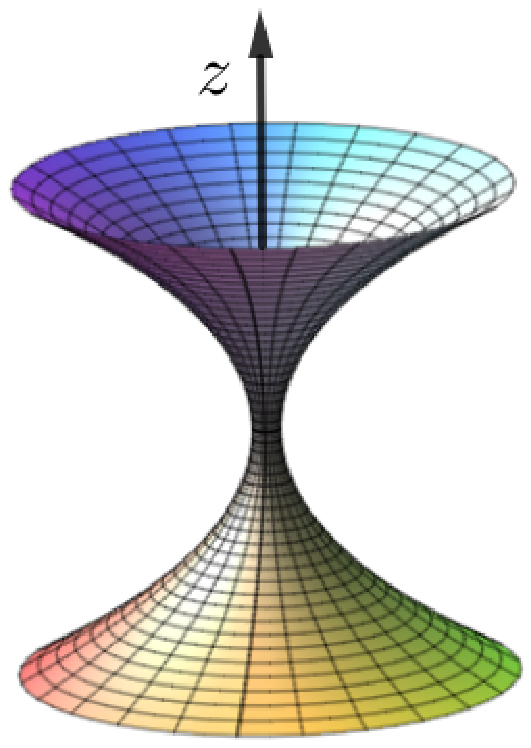}
 \caption{\label{kerr-wormhole} Plot showing the embedding diagrams of the Kerr-like wormholes. 
 (Left) Rotating Kerr-like wormhole: $z$ as a function of $r$ with $\phi$ direction for $M=1$ and $a=1$. 
 (Right) Slowly rotating Kerr-like wormhole: plot of $z$ as a function of $\rho $ with $\phi$ 
 direction for  $M=1.2$, $a=0.01$ and $\lambda^2=0.01$.} 
\end{figure*}

We commence with the Kerr-like wormhole spacetime which was obtained as a toy model by performing a 
modification on the Kerr metric  similar to  that of Damour and Solodukhin  for a Schwarzschild black 
hole \cite{Damour:2007ap}. This spacetime class was introduced by Bueno \emph{et al.} 
\cite{Bueno:2017hyj} and the metric in Boyer-Lindquist coordinates ($t, r, \theta, \phi$) is given by 
\begin{eqnarray}\label{metric}
ds^2 & = & -\left(1-\frac{2Mr}{\Sigma}\right)dt^2 -\frac{4Mar\sin^2 \theta}{\Sigma} dt d\phi 
+\frac{\Sigma}{\hat{\Delta}}dr^2   \nonumber \\ &+& \Sigma d \theta^2 
+ \left(r^2+ a^2 +\frac{2Ma^2r \sin^2 \theta}{\Sigma} \right) \sin^2 \theta d\phi^2,
\end{eqnarray}
where $\Sigma$ and $\hat{\Delta}$ are expressed as follows
\begin{eqnarray}\label{root}
\Sigma = r^2 + a^2 \cos^2\theta, \quad \hat{\Delta}=r^2 + a^2 - 2 M(1+\lambda^2)r.
\end{eqnarray}
This spacetime contains a family of parameters where $a$ and $M$, respectively corresponds to the spin 
and the mass of the wormhole and $\lambda^2$ is the deviation parameter which accounts for the deviation 
from the Kerr spacetime. In other words, for any non vanishing $\lambda^2$  the new metric differs from 
the Kerr metric. The Kerr spacetime may be recovered  when $\lambda^2 =0$. The throat of the Kerr-like 
wormhole (\ref{root}) can be evaluated by equating the $\hat{\Delta}$ to zero, which gives
\begin{equation}
r_{+} = M (1+\lambda^2) + \sqrt{M^2(1+\lambda^2)^2 -a^2}.
\end{equation}
This expression represents a special surface or region that connects two different asymptotically flat 
regions. Now we can construct the embedding diagrams to represent a Kerr-like wormhole and extract some 
useful information by considering an equatorial slice, $\theta  = \pi/2$ and a fixed moment of time, $t =$ 
a constant. The metric has the form 
\begin{equation}\label{em1}
ds^2=\frac{dr^2}{1-\frac{b(r)}{r}}+R^2 d \phi^2,
\end{equation}
where 
\begin{equation}\label{em2}
R^2=r^2+a^2+\frac{2Ma^2}{r},~~\text{and}~~b(r)=2M(1+\lambda^2)-\frac{a^2}{r}.
\end{equation}
We embed the metric (\ref{em1}) into three-dimensional Euclidean space to visualize this slice and the 
spacetime can be written in cylindrical coordinates as
\begin{equation}\label{em3}
ds^2=dz^2+dR^2 + R^2 d\phi^2=\left[\left(\frac{dR}{dr} \right)^2+\left(\frac{dz}{dr} \right)^2\right]dr^2+R^2 d \phi^2.
\end{equation}
The combination of (\ref{em1}) and (\ref{em3}) generate the equation for the embedding surface which is given by
\begin{equation}\label{em4}
\frac{dz}{dr}=\pm \sqrt{\frac{r}{r-b(r)}-\left(\frac{dR}{dr} \right)^2}.
\end{equation}
Now, plugging the value of (\ref{em2}) into (\ref{em4})  the above relation assumes the form
\begin{equation}
\frac{dz}{dr}=\pm \sqrt{\frac{M\left[2r^7\lambda_1 +2 a^2 r^3(2r^2+ a^2) 
-4Ma^2 \lambda_1 r^4 -Ma^4(r^2+a^2)+2Ma^4 \lambda_1 r\right]}
{r^3(r^3+a^2r+2Ma^2)\left(r^2+a^2-2M\lambda_1 r\right)} },\label{2.9}
\end{equation}
where we have put $\lambda_1 = 1 + \lambda^2$ for convenience. Note that the integration of (\ref{2.9}) 
cannot be accomplished analytically. Invoking numerical techniques allows us to  illustrate the wormhole 
shape given in Fig.~\ref{kerr-wormhole}.
In the limit $a\rightarrow 0$, we obtain
\begin{equation}
z= \pm \sqrt{8M(1+\lambda^2) (r-2M-2M\lambda^2)},
\end{equation}
for the integral of (\ref{2.9}). In addition, on setting the deviation parameter to zero, i.e.,  
$\lambda^2=0$, the Schwarzschild solution  
\begin{equation}
z= \pm \sqrt{8M (r-2M)}.
\end{equation}
is regained. The problem may be simplified by considering a slowly rotating Kerr-like wormhole solution. 
For this purpose we are interested in approximate solutions by considering a series expansion, say, around 
$a$ and $\lambda^2$ in Eq.~\eqref{2.9}. The solution can be approximated as follows
\begin{equation}\notag
z= \pm \Big\{\sqrt{8M (r-2M)}-\frac{\lambda^2 \sqrt{2M} }{4 M^{3/2} r(r-2M)^{3/2}}
\left[\sqrt{2}\arctan \left(\frac{r-2M}{\sqrt{M}} \right)(r-2M)^{3/2}r a^2+ \chi  \right]
\end{equation}
\begin{equation}
+\frac{a^2 \sqrt{2M} }{16 r^2 M^{3/2} (r-2M)^{1/2}} \left[3\sqrt{2}\arctan \left(\frac{r-2M}{\sqrt{M}}  \right)(r-2M)^{1/2}r^2+\zeta  \right]\Big\},
\end{equation}
where 
\begin{equation}
\zeta =16M^{5/2}-4 M^{3/2}r+6 r^2 M^{1/2},
\end{equation}
and 
\begin{equation}
\chi = 24 M^{5/2}r^2-32M^{7/2}r+4M^{5/2}a^2-4 M^{3/2}r^3-8 a^2 M^{3/2}r+2 r^2 a^2 \sqrt{M}.
\end{equation}
Finally, let us introduce a new variable $\rho$, via 
\begin{equation}\label{emb}
r=2M(1+\lambda^2)+\frac{\lambda^2-1}{2M}a^2+\frac{\rho^2}{2M}.
\end{equation}
The graphical representation of (\ref{emb}) with the $\phi$ direction can be seen from 
Fig.~\ref{kerr-wormhole}, which shows the visual image of the  slowly rotating Kerr-like wormhole.  

\subsection{ADM mass}
Now let us compute the ADM mass for the Kerr-like wormhole. We compute it for asymptotically flat 
spacetime and it will be similar to that of axisymmetric spacetime. We consider the asymptotic flat 
spacetime
\begin{eqnarray}
ds^2_{\Sigma} = \psi(r)dr^2+r^2 \chi(r)\left(d\theta^2+\sin^2\theta d\varphi^2\right),
\end{eqnarray}
where we have identified
\begin{equation}
    \psi(r)=\frac{1}{1-\frac{2M(1+\lambda^2)}{r}},\quad \text{and} \quad \chi(r)=1.
\end{equation}
In order to compute the ADM mass, we use the approach which used in \cite{Shaikh:2018kfv},
\begin{equation}\label{for}
    M_{ADM}=\lim_{r\to \infty} \frac{1}{2}\left[-r^2 \chi'+r(\psi -\chi) \right].
\end{equation}
On substituting the values in (\ref{for}) and after computing the limit we get the ADM mass for the 
wormhole,
\begin{equation}\label{ADM}
    M_{ADM}=M(1+\lambda^2).
\end{equation}
Note that this is the mass of the Kerr-like wormhole as seen by an observer located at the 
asymptotic spatial infinity.  We will use this mass to explain the physical processes throughout the paper.

\section{Geodesics in Kerr-like wormholes}
\label{geodesics}
The spacetime of rotating Kerr-like wormholes is generally characterized by the four constants of 
motion, i.e., the Lagrangian $\mathcal{L}$, the energy $E$, the angular momentum $L_{z}$ and the Carter 
constant $\mathcal{K}$. The geodesic equations of a test particle having rest mass $m_0$ can be derived 
by using these conserved quantities and the Hamilton-Jacobi equation. We can derive the geodesic equations 
for $t$ and $\phi$ coordinates with the help of conserved quantities, $E=-p_t$ and $L_z=p_{\phi}$, which 
turns out
\begin{eqnarray}\label{u}
\Sigma \frac{d t}{d \sigma} &=&  -a \left(aE \sin^2 \theta - L_{z}\right) 
+ \frac{(r^2 + a^2) \mathcal{P}}{r^2 + a^2 - 2 M r},
\nonumber \\
\Sigma \frac{d \phi}{d \sigma} &=& -\left(aE - L_{z}\csc^2 \theta \right) 
+ \frac{a \mathcal{P}}{r^2 + a^2 - 2 M r},
\end{eqnarray}
where $\mathcal{P} = (r^2+a^2)E-aL_{z}$. The other geodesic equations, i.e., for $r$ and $\theta$ 
coordinates can be computed by using the Hamilton-Jacobi method. Since the geodesics of the test particle 
in the background of Kerr-like wormholes spacetime (\ref{metric}) satisfy the Hamilton-Jacobi equation 
\cite{Chandrasekhar:1992}, which is given by
\begin{equation}\label{HamJac}
\frac{\partial S}{\partial \sigma} = -\frac{1}{2} g^{\mu \nu} 
\frac{\partial S}{\partial x^{\mu}} \frac{\partial S}{\partial x^{\nu}},
\end{equation}
where $\sigma$ is the affine parameter and $S$ is the Jacobian action with the following separable ansatz:
\begin{equation}\label{ansatz}
S = \frac{1}{2} m_0^2 \sigma -Et +S_r(r) +S_{\theta}(\theta) +L_z \phi.
\end{equation}
Here $S_r$ and $S_{\theta}$ are  functions of $r$ and $\theta$, respectively. On substituting 
Eq.~(\ref{ansatz}) into Eq.~(\ref{HamJac}) and after some straightforward computations we obtain the 
following forms of the geodesic equations
\begin{eqnarray}\label{u1}
\Sigma \frac{d r}{d \sigma} &=& \pm  \sqrt{\mathcal{R}},
\nonumber \\
\Sigma \frac{d \theta}{d \sigma} &=& \pm  \sqrt{\Theta},
\end{eqnarray}
where the terms $\mathcal{R}$ and $\Theta$ in (\ref{u1}) are expressed as follows
\begin{eqnarray}\label{quant}
\mathcal{R} &=& \frac{\left[r^2 + a^2- 2 M (1+\lambda^2) r \right]
\lbrace \mathcal{P}^2 -(r^2 + a^2- 2 M r) 
\left[\mathcal{K}+ (L_{z}-a E)^2 +m_0^2r^2 \right] \rbrace }{r^2 + a^2- 2 M r}, 
\nonumber \\
\Theta &=& \mathcal{K} +\cos^2 \theta \left(a^2E^2-L_{z}^2\csc^2 \theta -m_0^2 a^2 \right).
\end{eqnarray}
These geodesic equations determine the trajectories of a test particle in background of the Kerr-like 
wormholes spacetime. Physically, the quantity $E$ is the energy required for a distant observer to place 
the test particle in orbit around the Kerr-like wormhole.

\section{Spherical photon orbits around Kerr-like wormholes}
\label{orbit}
Consider the  case of photons released from a bright source and which are moving towards the wormhole 
that is placed between the observer and the light source. The possible trajectories of the photons 
around the wormhole are: (i) falling into the wormholes, (ii) scattered away from the wormholes to 
infinity, and (iii) critical geodesics which separate the first two sets, also known as unstable 
spherical orbits. It turns out that in the observer's sky, the plunged photon geodesics form dark spots 
whereas the scattered photon geodesics form bright spots. The critical photon geodesic trajectories form 
a dark region in the observer's sky in the presence of a bright background which is referred to as the 
\emph{wormhole hole shadow}.

Our principal objective is to evaluate these critical geodesics or unstable spherical orbits. In order 
to obtain the boundary of the wormhole shadow, it is necessary to work out the radial motion of photons 
around the wormholes. Here, we consider a photon as a massless test particle, therefore $m_0=0$. The 
radial geodesic equation can be rewritten as 
\begin{equation}\label{rad}
 \left(\Sigma \frac{dr}{d\sigma}\right)^2 + V_{eff} =0,
\end{equation}
where $V_{eff}$ represents the effective potential that describes the trajectories of the photons around 
the wormhole. However, the different bounds of the effective potential determine the various 
trajectories depending on the magnitude of the angular momentum. Now we express the effective potential 
of the photons in terms of two independent impact parameters such that $\xi=L_{z}/E$ and 
$\eta=\mathcal{K}/E^2$. The effective potential can be written as follows
\begin{eqnarray}
\label{Veff}
V_{eff} &=& -\frac{E^2 \left[r^2 + a^2- 2 M(1+\lambda^2)r\right]}{r^2 + a^2 -2 Mr} 
\Big[(r^2+a^2-a\xi)^2 \nonumber\\ &&
-(r^2 + a^2- 2 M r) \left[\eta + (\xi -a)^2 \right] \Big].
\end{eqnarray}
It is easy to observe that the effective potential has a dependency of the deviation parameter $\lambda^2$ 
that affects the trajectories of photons.
\begin{figure*}
\includegraphics[scale=0.6]{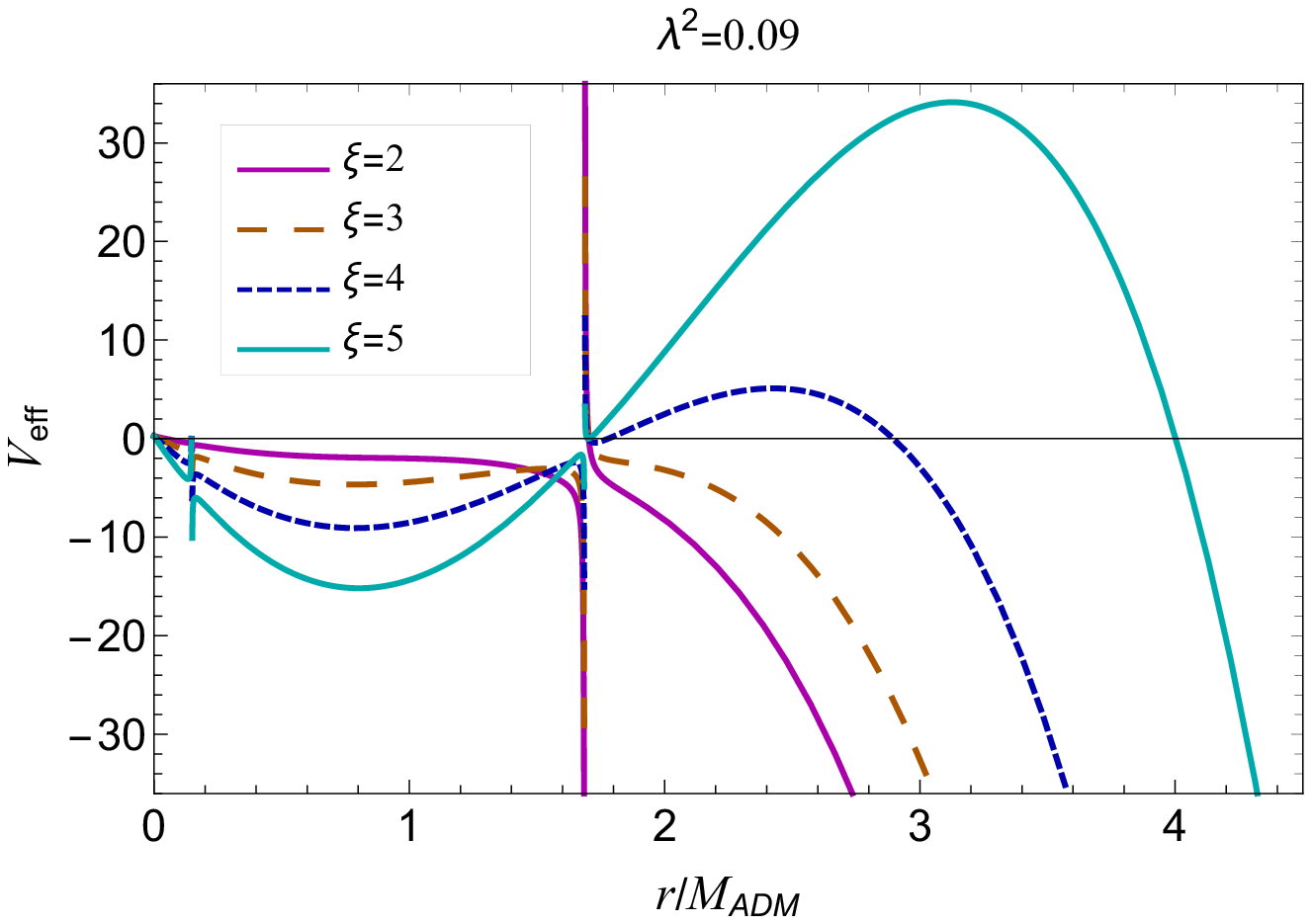}
\includegraphics[scale=0.6]{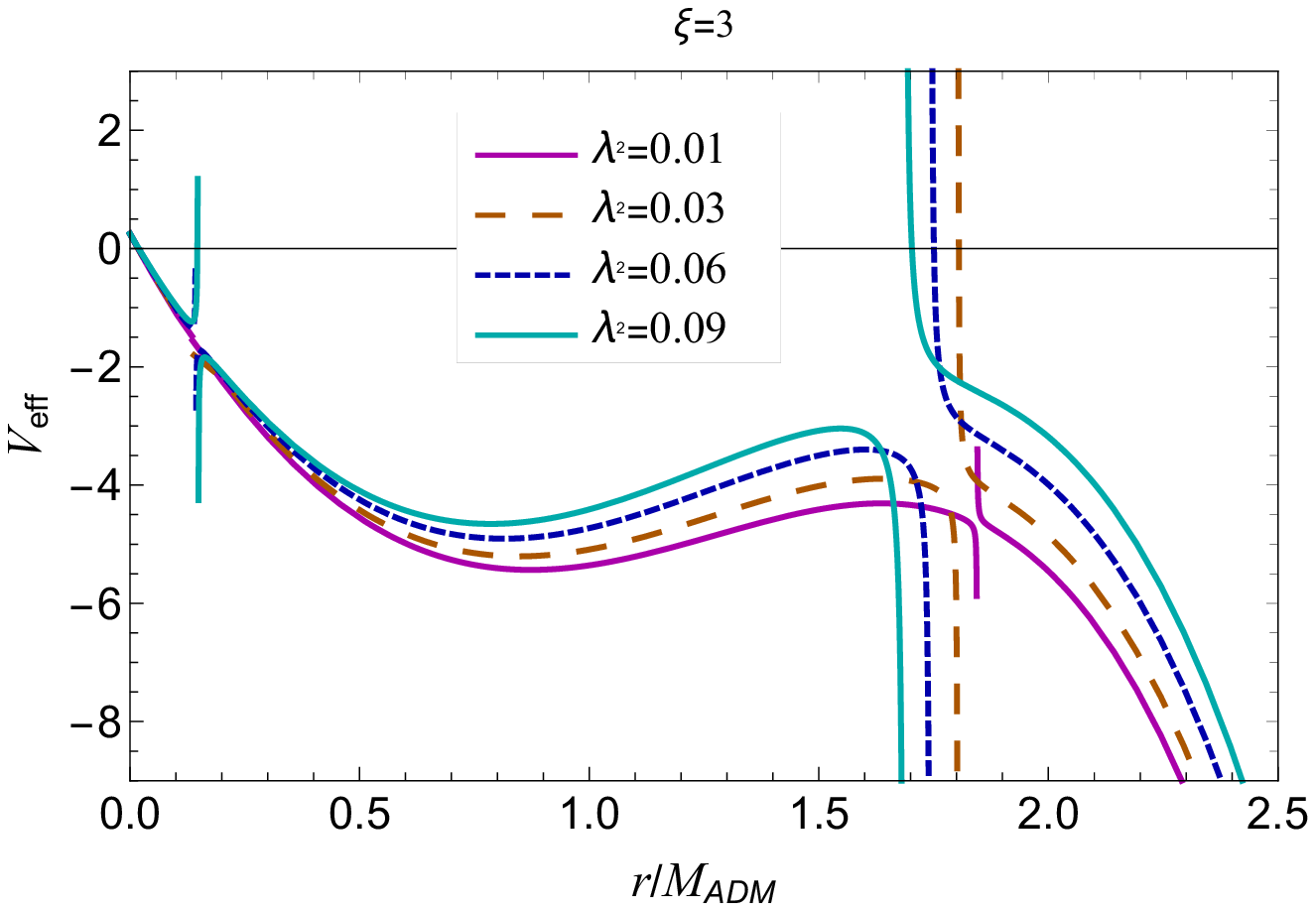}
\caption{\label{eff} Plot illustrating the nature of effective potential in Kerr-like wormholes by 
varying the impact parameter $\xi$ and the deviation parameter $\lambda^2$.}
\end{figure*}
We plot effective potential with radius for the Kerr-like wormhole in Fig.~\ref{eff} by varying the 
parameter $\xi$ as well as the deviation parameter $\lambda^2$. We use the ADM mass ($M_{ADM}$) to plot 
the effective potential of the Kerr-like wormholes. We find that the effective potential has different 
potential barriers according to the variation of parameters which increase with parameter $\xi$ as well 
as with the deviation parameter $\lambda^2$. On the other hand, the boundary of the shadow can be 
determined by the unstable spherical photon orbits corresponding to the highest maximum value (supremum) 
of the effective potential $V_{eff}$. The following standard conditions can be used to determine the 
unstable spherical photon orbits
\begin{equation}\label{condition1}
V_{eff} = 0, \quad  V_{eff}'=0,
\end{equation}
where prime ($'$) represents derivatives with respect to $r$. In terms of $\mathcal{R}$, the unstable 
spherical photon orbits conditions transform as follows
\begin{equation}\label{condition}
\mathcal{R} = 0, \quad  \mathcal{R}'=0.
\end{equation}
Now on using Eqs.~(\ref{quant}) and (\ref{condition}), we immediately obtain the expressions of the 
impact parameters
\begin{eqnarray}\label{imptp}
\xi &=&\frac{a^2(r+M)+r^2 (r-3 M)}{a(M-r)},
\nonumber \\ 
\eta &=& \frac{4M a^2 r^3-r^4 (r-3 M)^2}{a^2(M-r)^2},
\end{eqnarray}
where $r$ is the radius of unstable spherical photon orbits. Interestingly, Eq.~(\ref{imptp}) determines 
the critical locus of the impact parameters which can be considered as the set of unstable photon orbits. 
As it can be seen from Eq.~(\ref{imptp}) that the impact parameters have a dependency on the spin $a$ 
and the mass $M$. When we compare our computed expressions of the impact parameters with the Kerr 
spacetime, we find them indistinguishable. However, there is a dependency of deviation parameter 
$\lambda^2$ on the impact parameters through the ADM mass by using the relation, 
$M = M_{ADM}/(1 +\lambda^2)$. If we compute the impact parameters for $a = 0$ (nonrotating case), 
we obtain the following relation
\begin{equation}
    \eta = 27M^2 - \xi^2 \quad \text{with} \quad r = 3M.
\end{equation}
This result is similar to that of the Schwarzschild spacetime one. Although we are using the ADM 
mass which turns out that there is a dependency on deviation parameter $\lambda^2$. The apparent 
image or shadow of the wormhole can be obtained by looking for the unstable photon orbits which can 
be discussed more explicitly in the next section.

\section{Shadow of Kerr-like wormholes}
\label{shadow}
\begin{figure*}
  \includegraphics[scale=0.34]{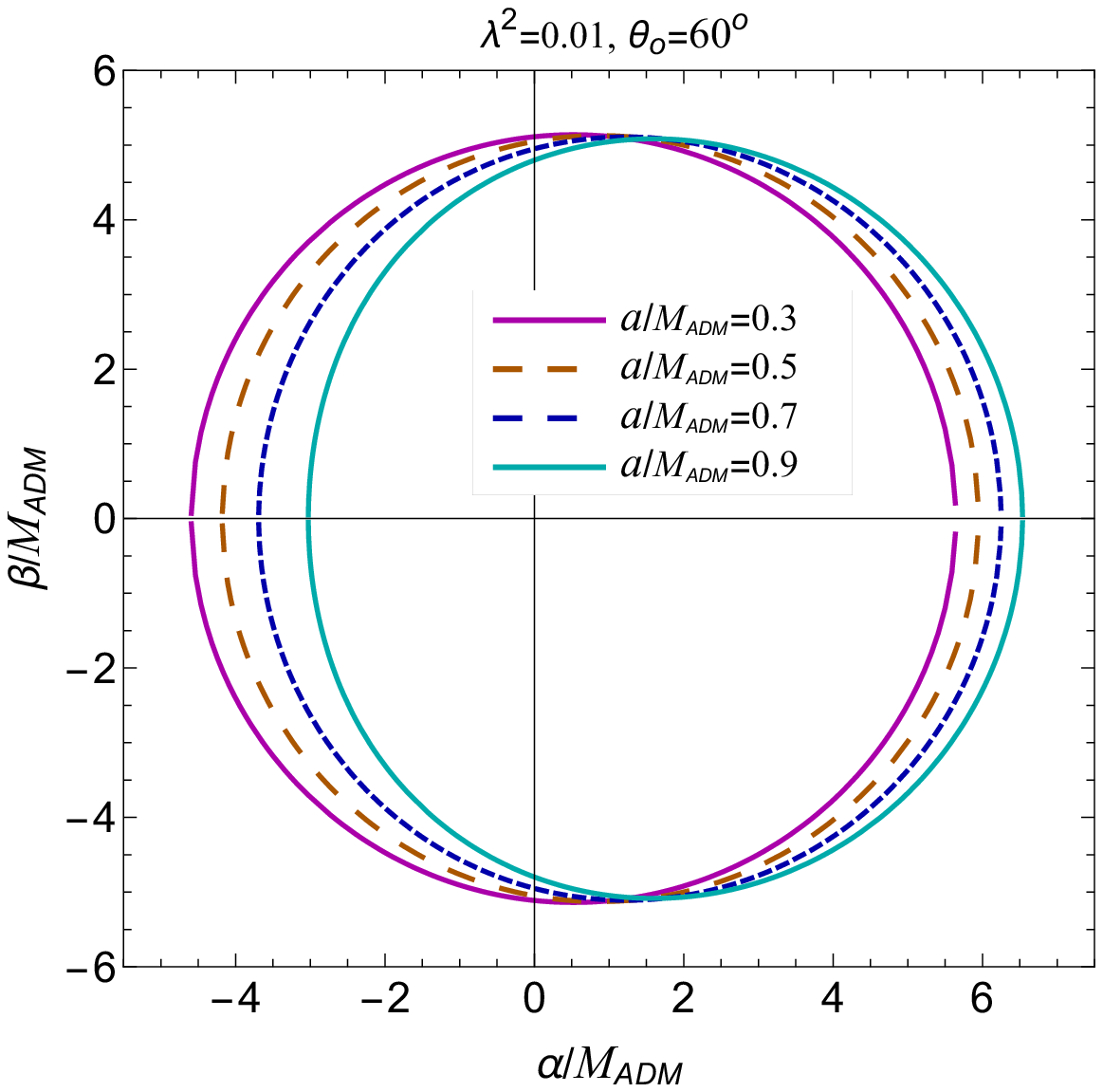}
  \includegraphics[scale=0.34]{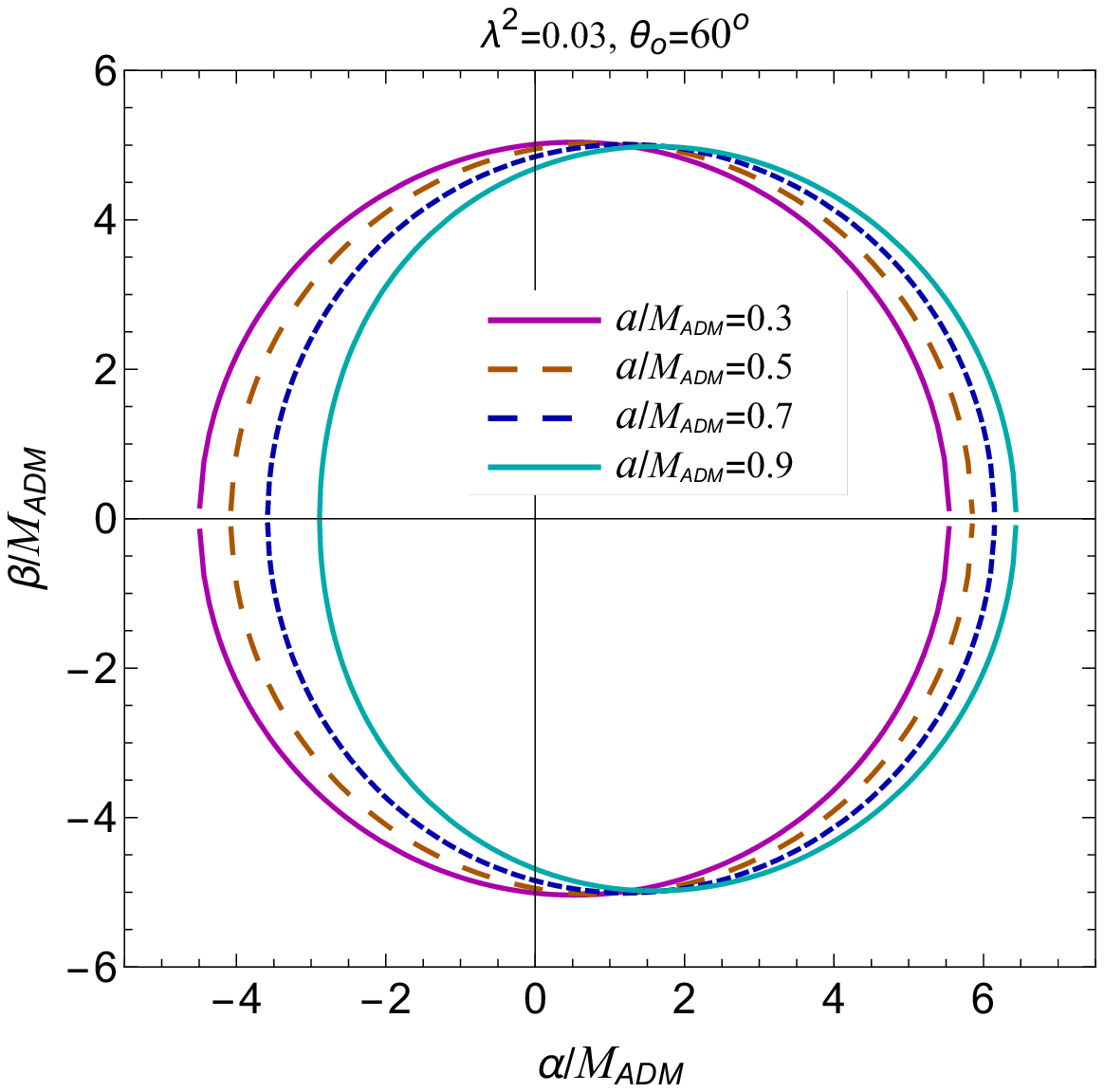}
  \includegraphics[scale=0.34]{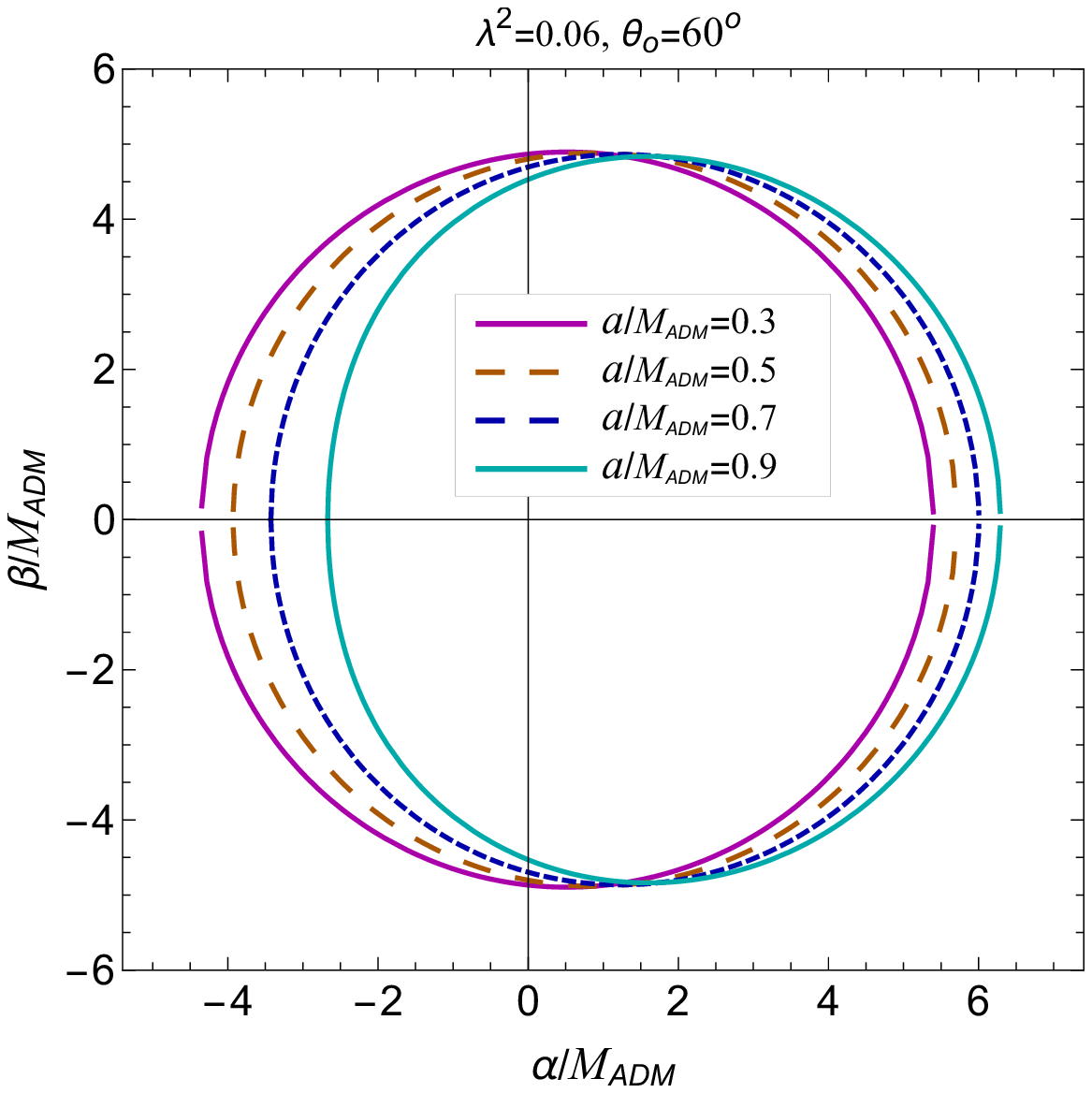}
  \includegraphics[scale=0.323]{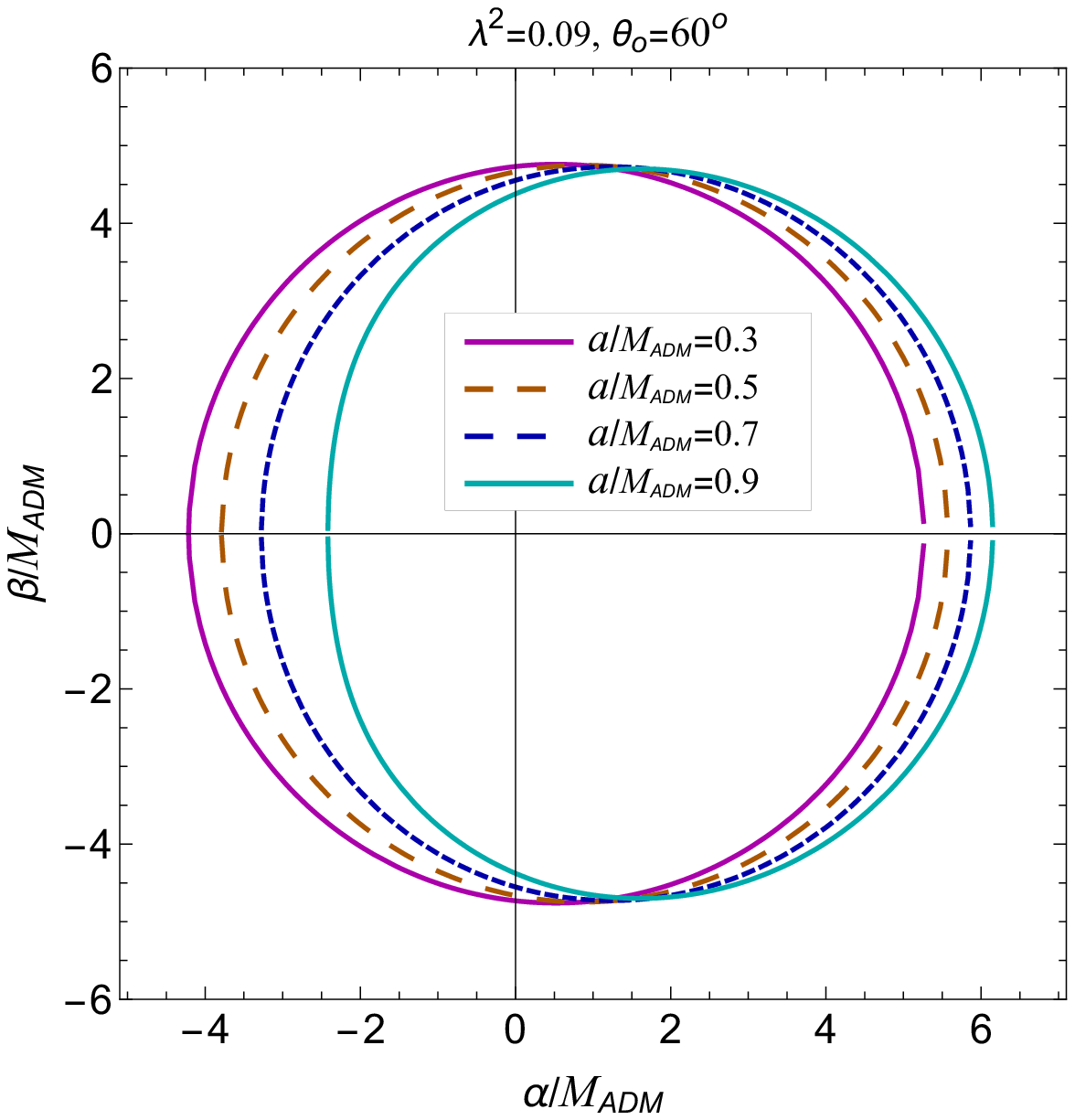}\\
  \includegraphics[scale=0.34]{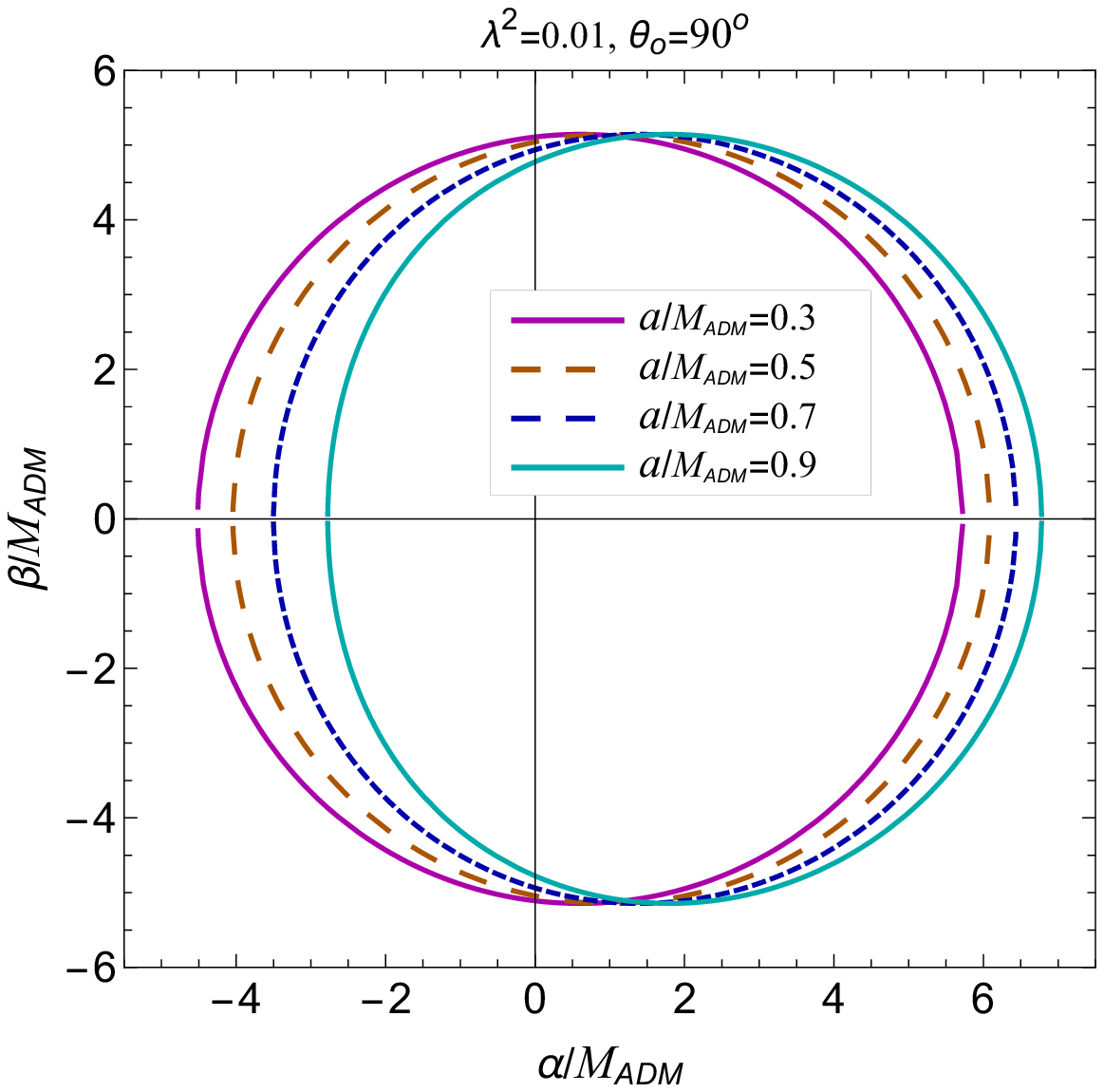}
  \includegraphics[scale=0.345]{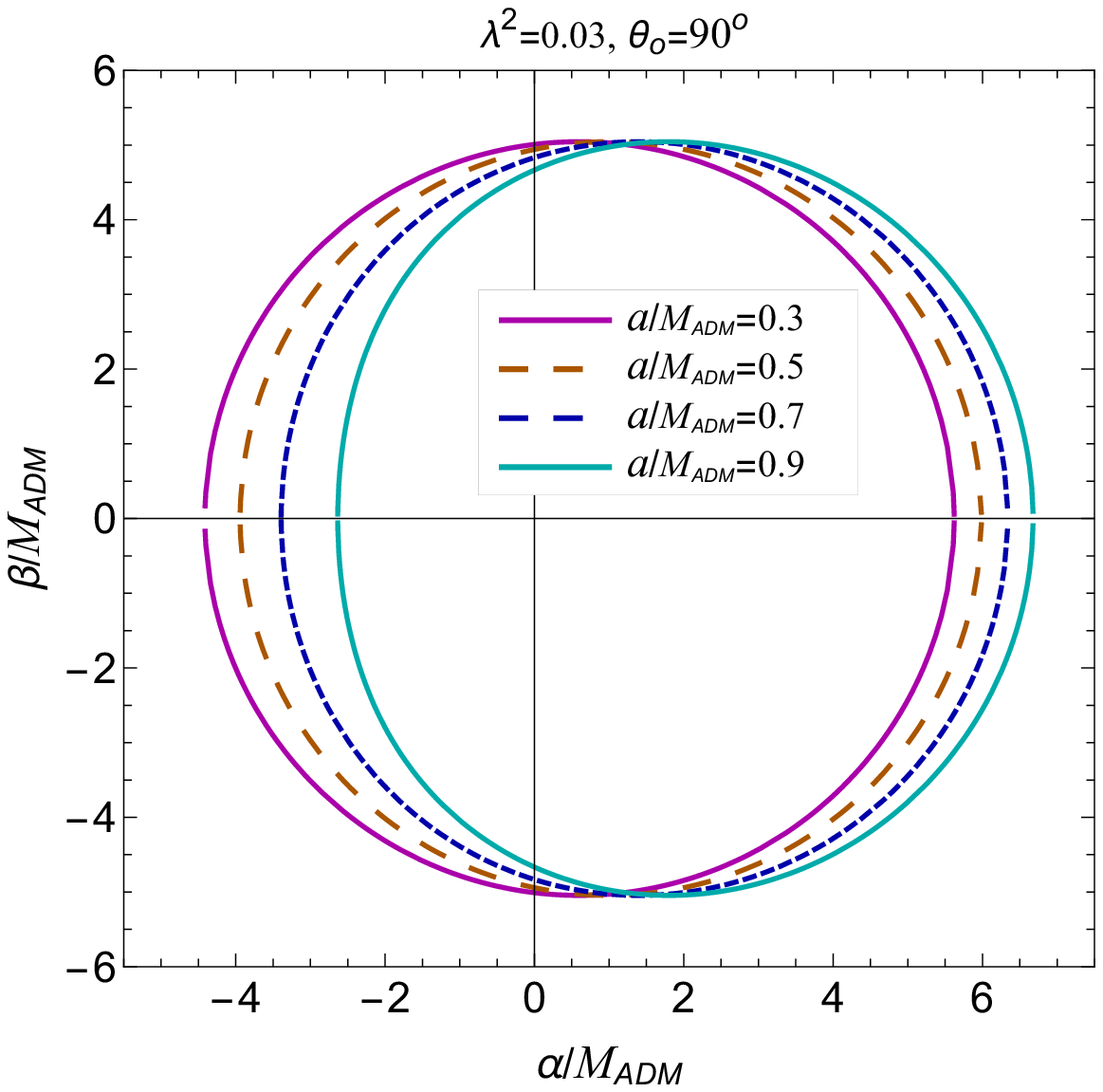}
  \includegraphics[scale=0.34]{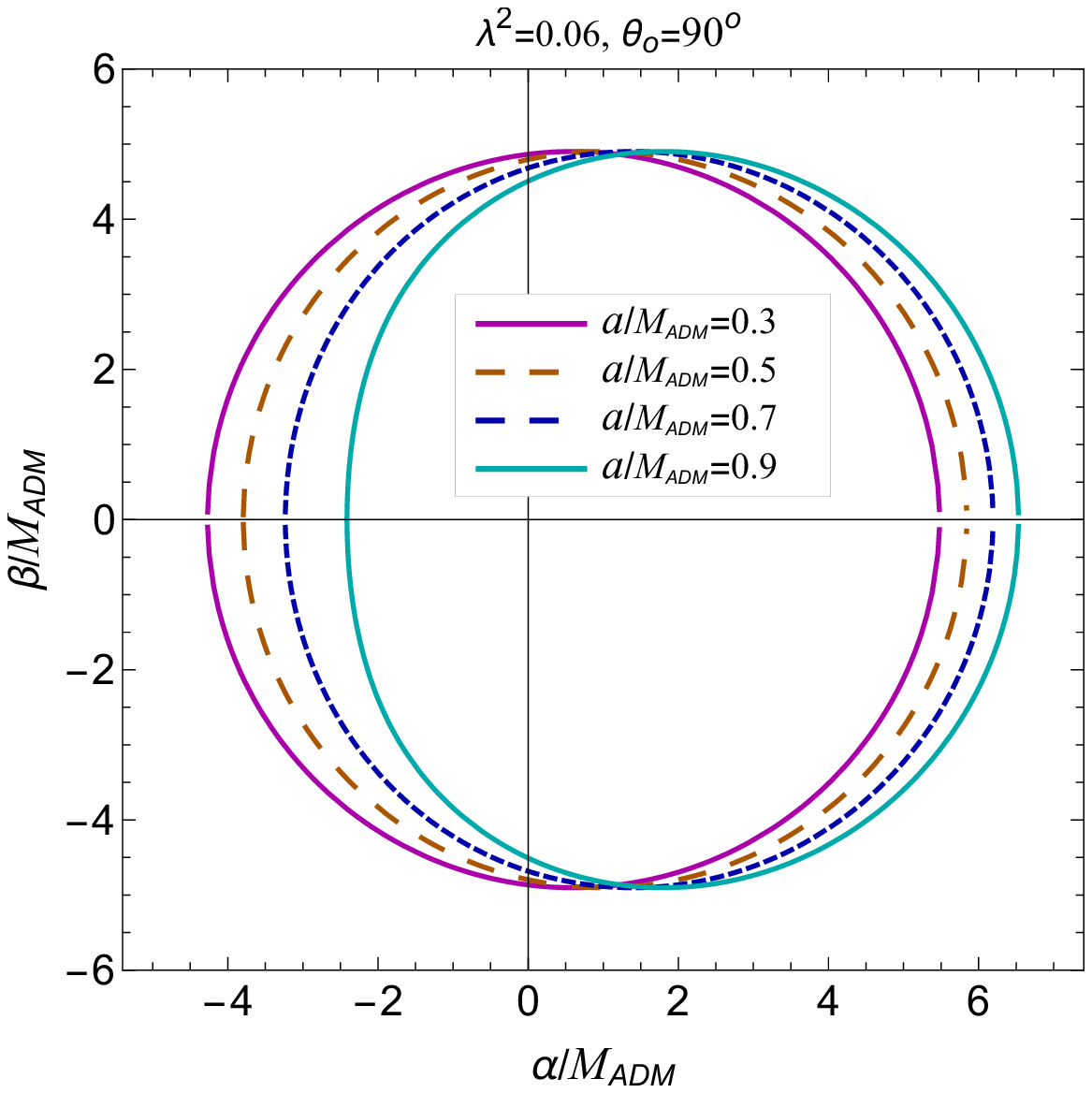}
  \includegraphics[scale=0.325]{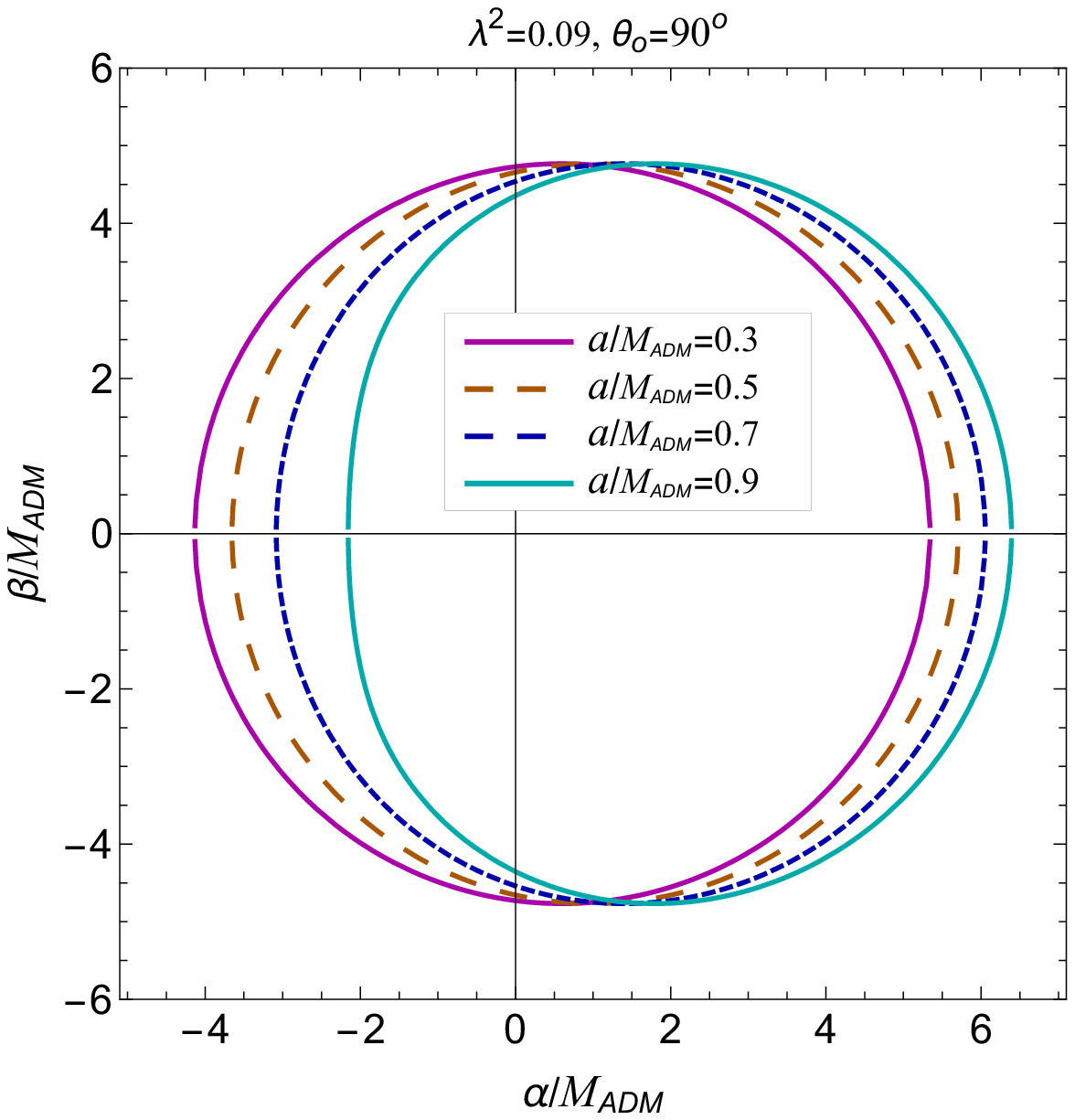}
 \caption{\label{shad} Plot illustrating the shadow images of the Kerr-like wormholes by varying 
 the spin $a$ for different values of deviation parameter $\lambda^2$ ($M_{ADM}=1$).}
\end{figure*}

\begin{figure*}
   \includegraphics[scale=0.34]{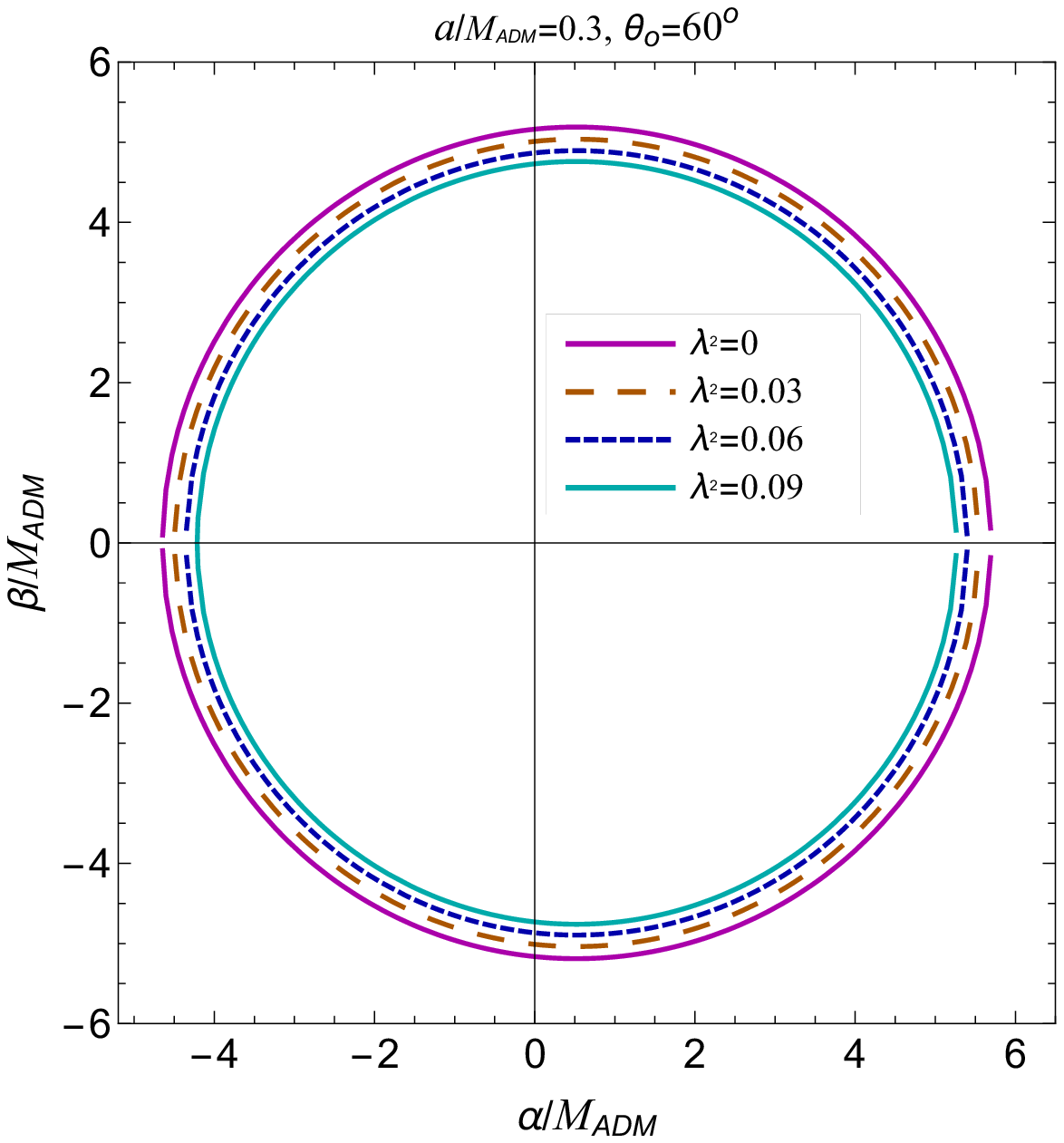}
   \includegraphics[scale=0.35]{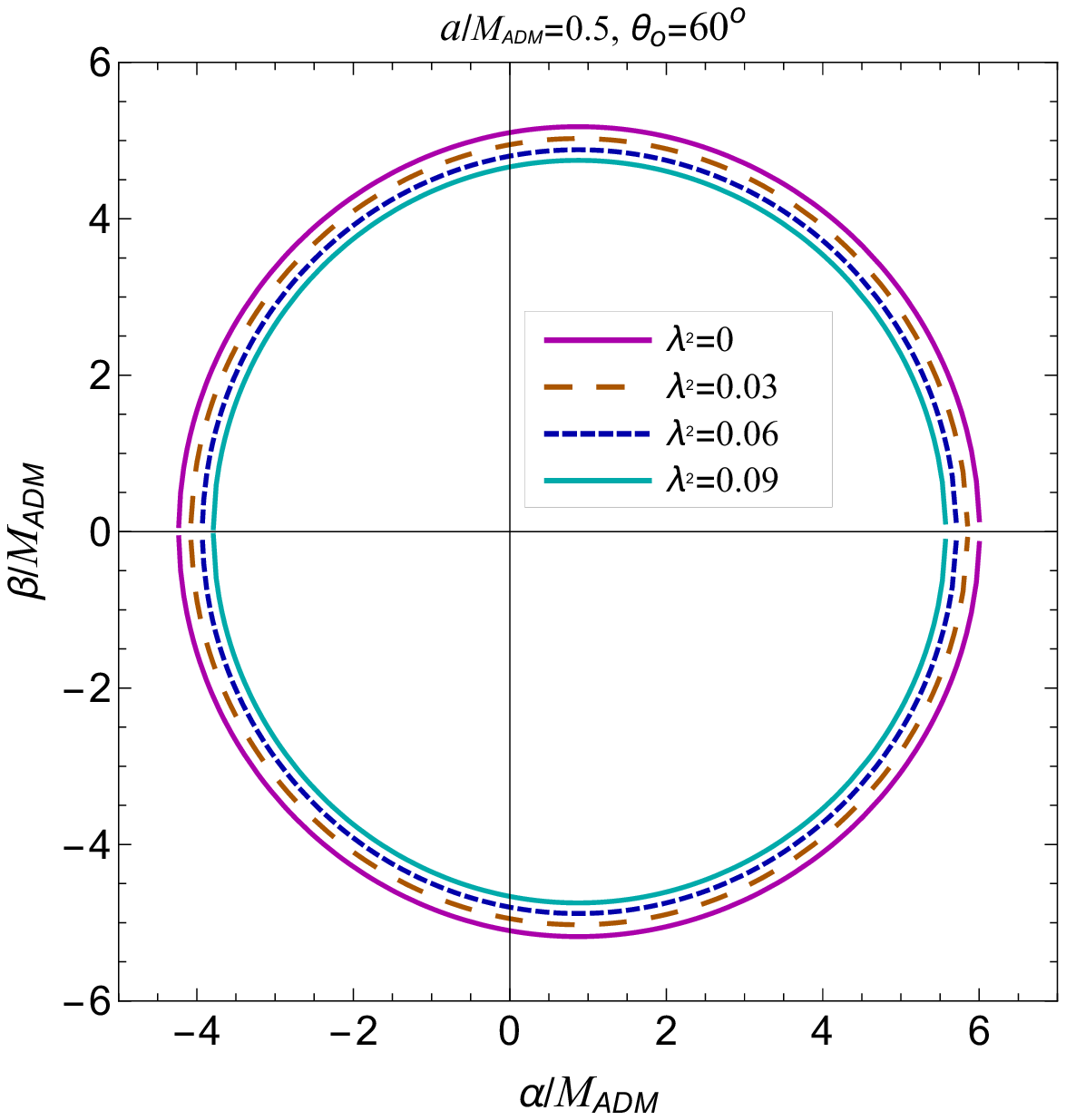}
   \includegraphics[scale=0.335]{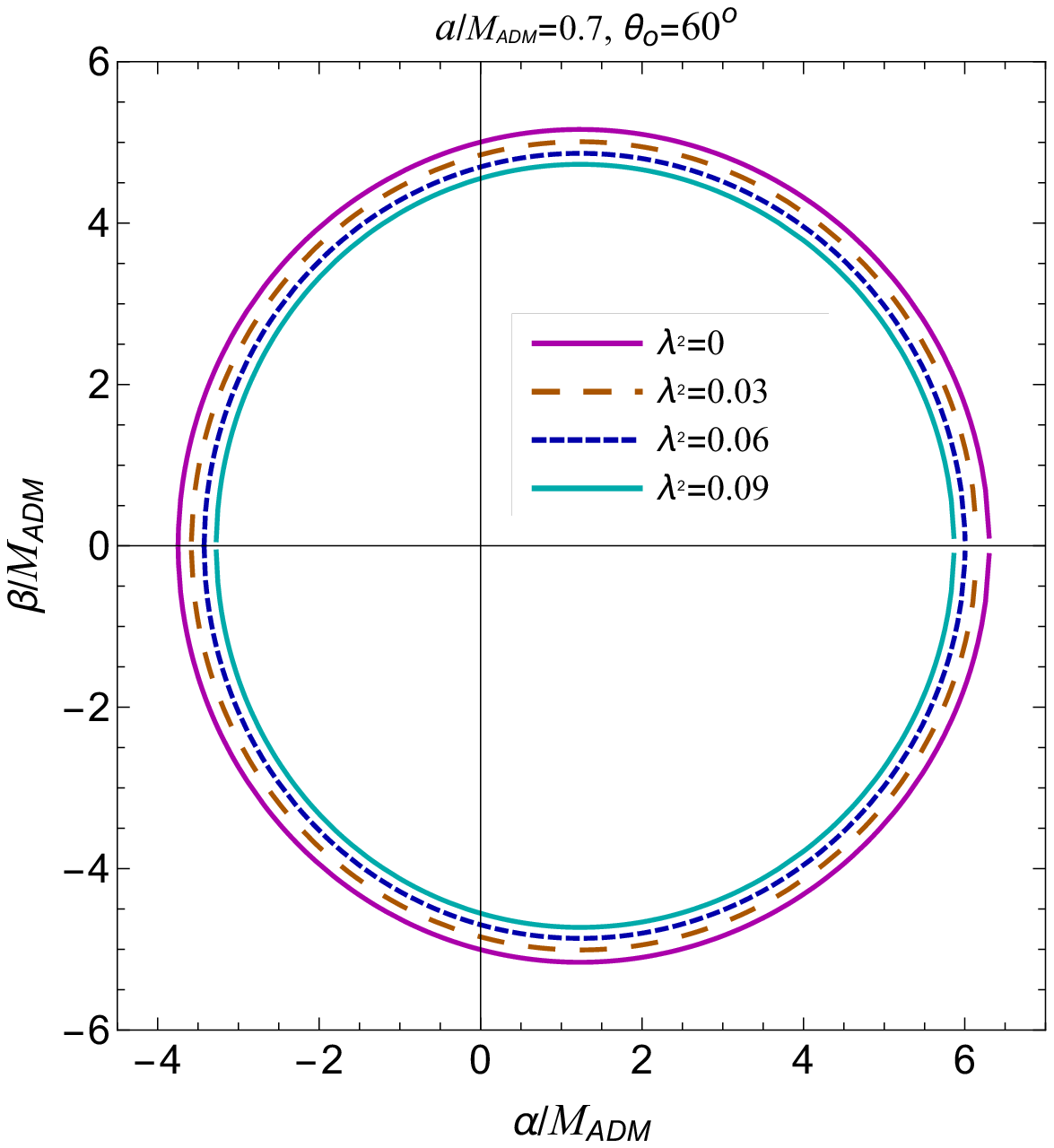}
   \includegraphics[scale=0.33]{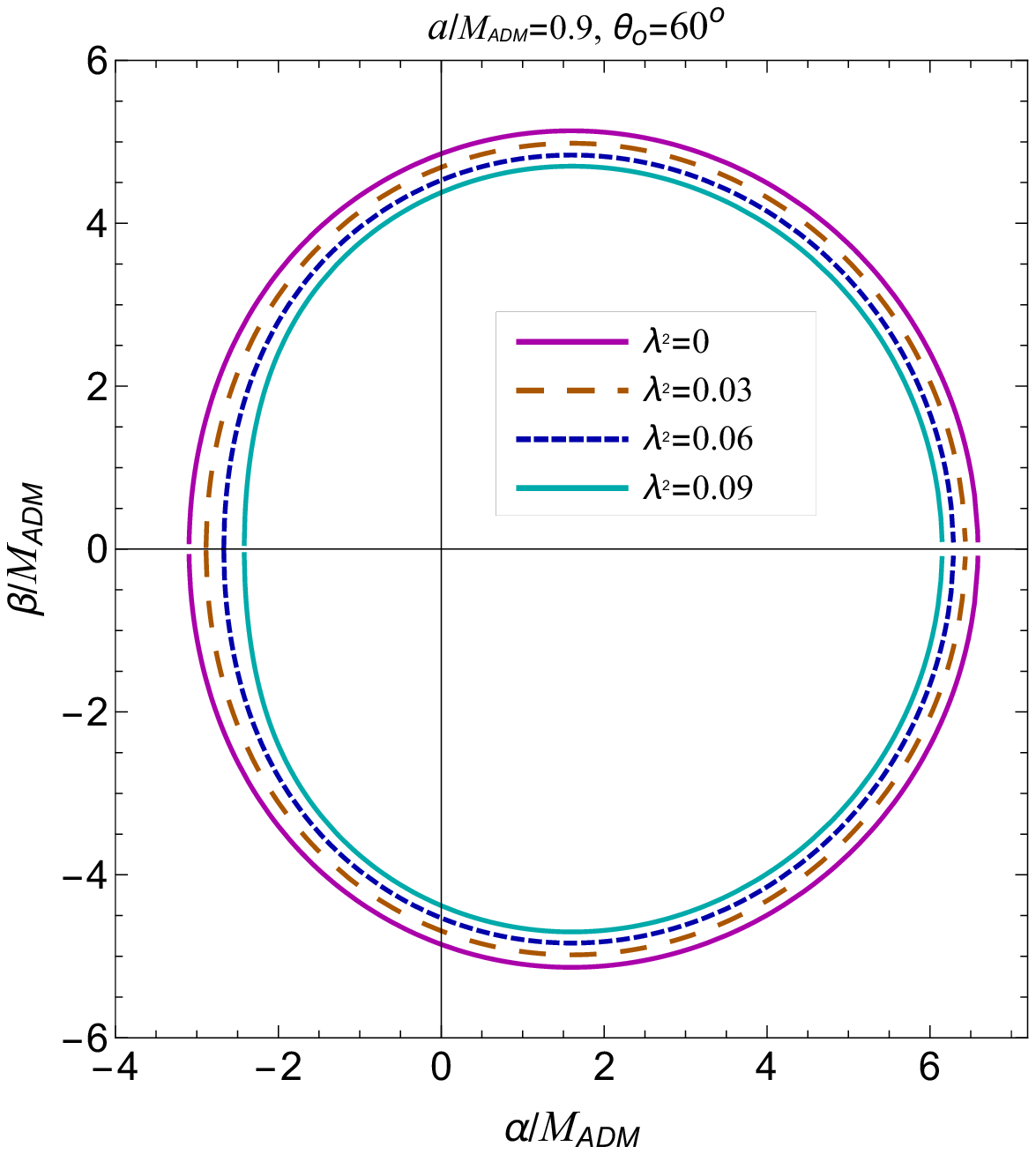}\\
   \includegraphics[scale=0.34]{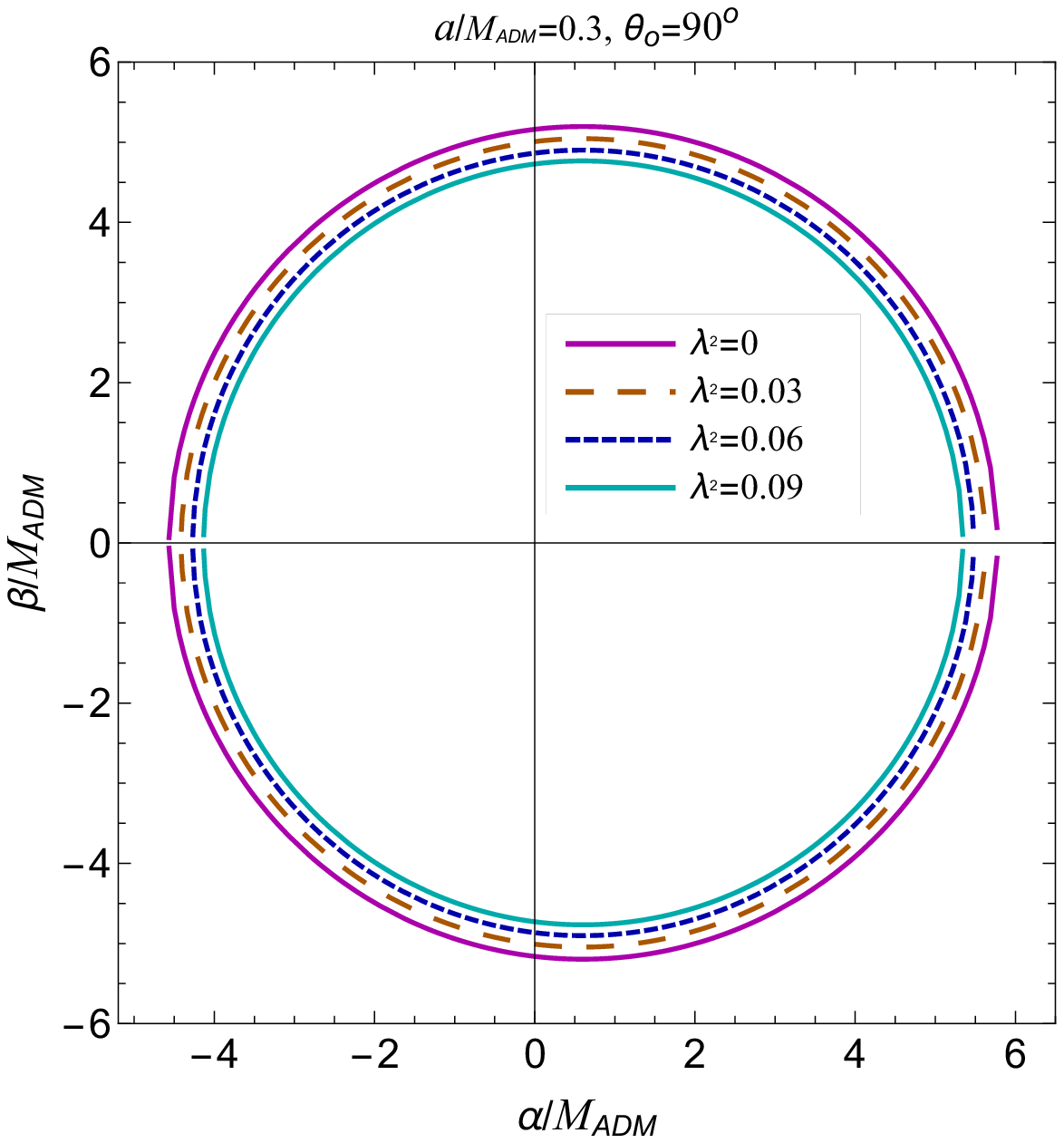}
   \includegraphics[scale=0.35]{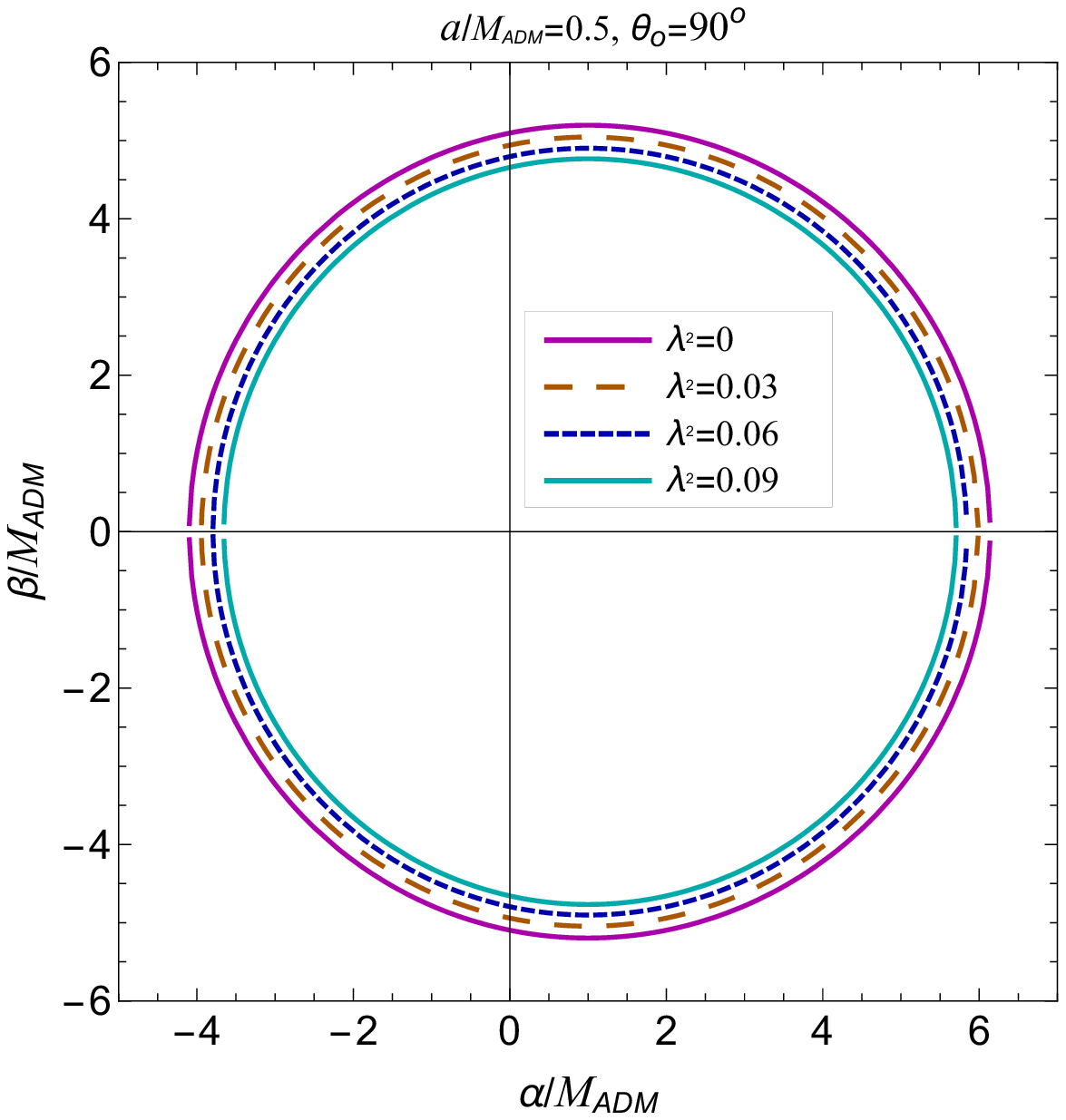}
   \includegraphics[scale=0.335]{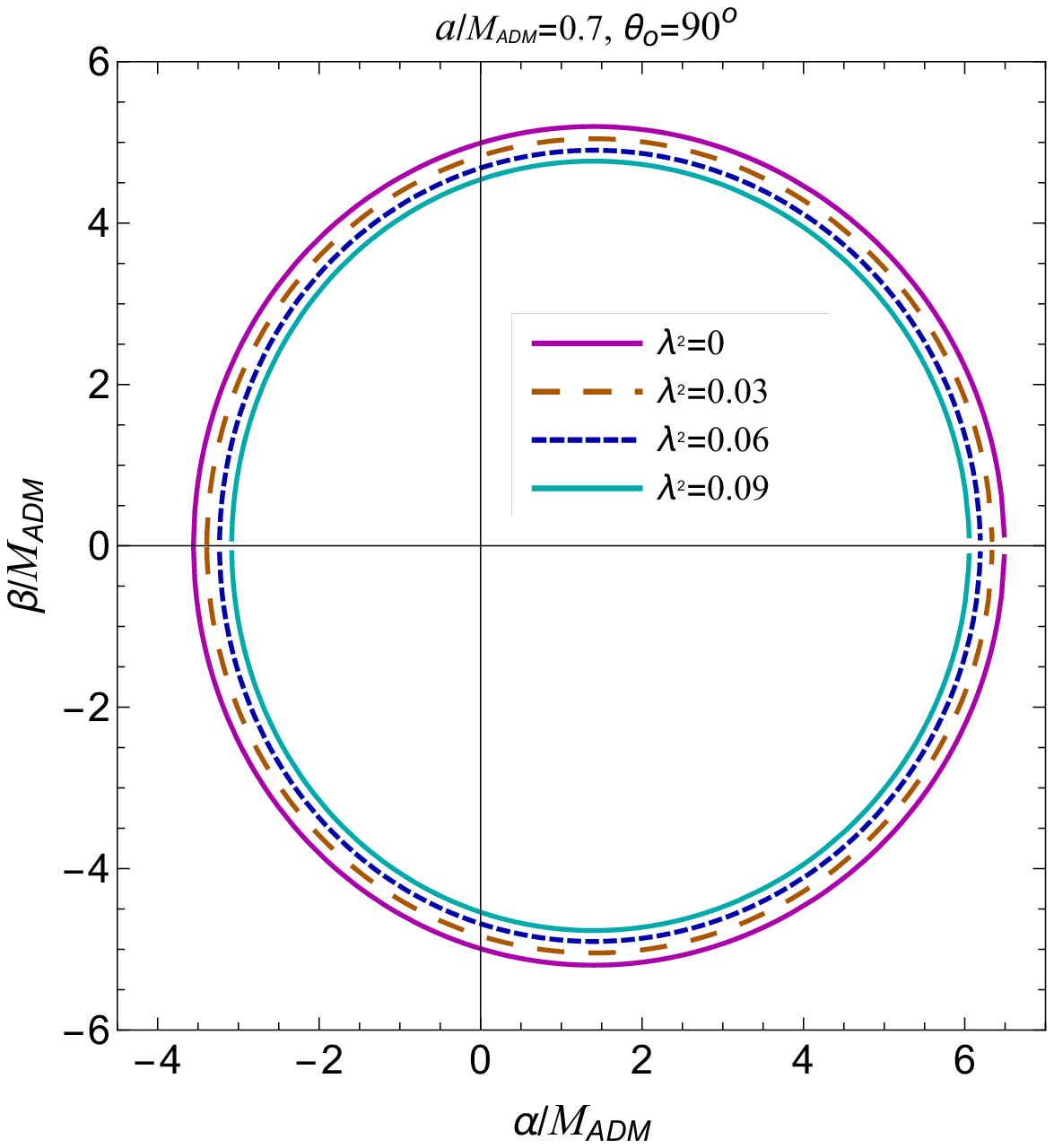}
   \includegraphics[scale=0.33]{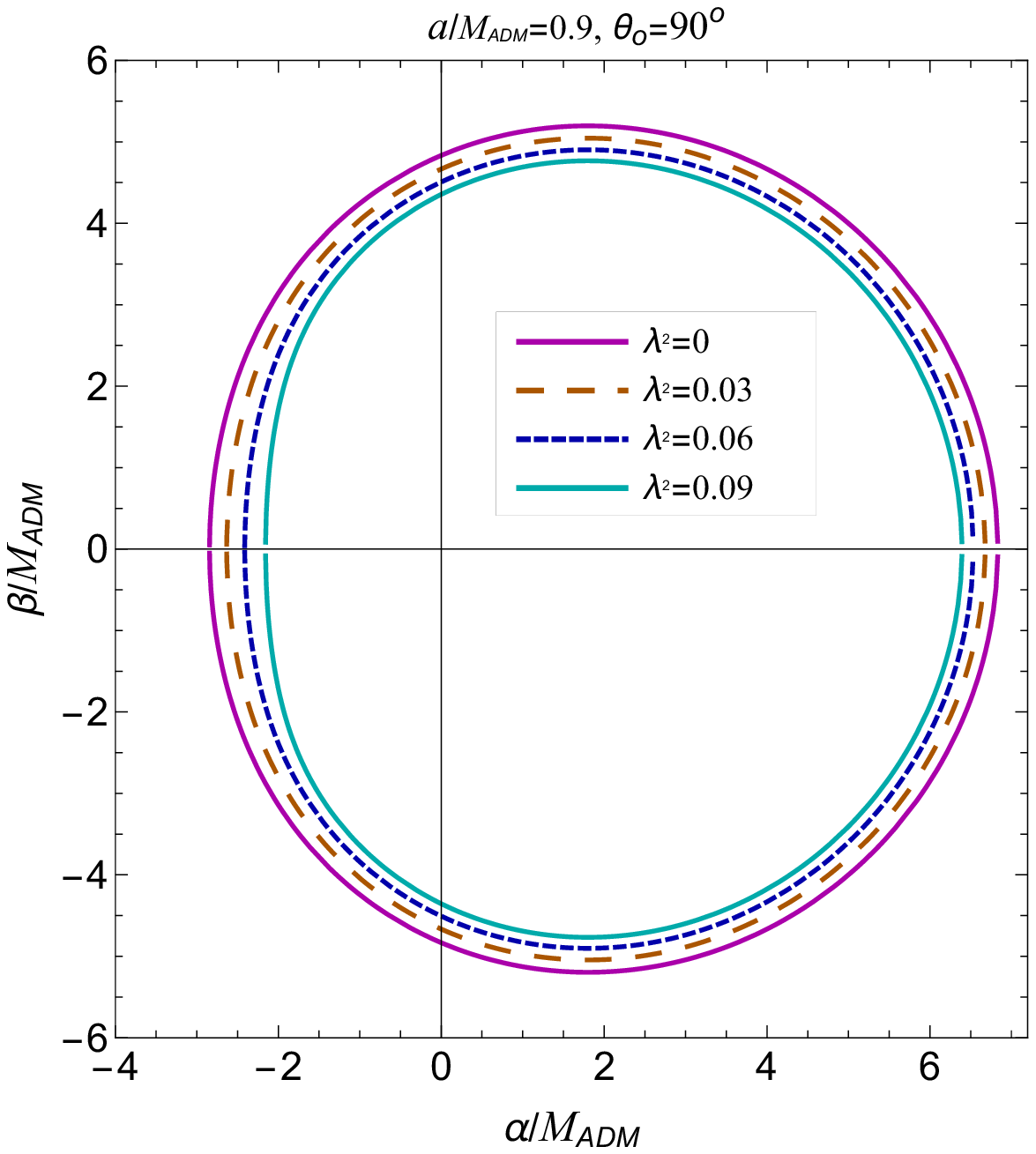}
 \caption{\label{shad-1} Plot illustrating the shadow images of the Kerr-like wormholes by varying 
 the deviation parameter $\lambda^2$. Here $\lambda^2=0$ corresponds to the Kerr black hole case 
 ($M_{ADM}$ is set equal to $1$ for both of the spacetimes).}
\end{figure*}

In this section, our main goal is to construct and characterize the shadow of the Kerr-like wormholes. 
We turn our attention towards the observer's sky to detect the optical images cast by the Kerr-like 
wormholes. In preparation for the investigation of the shadow,  consider a plane passing through the 
center of the wormhole and it's normal joining the center of the wormhole and the line of sight of an 
observer. We introduce new coordinates ($\alpha,\beta$) \cite{Bardeen:1973gb}, widely known as 
celestial coordinates that span a two-dimensional plane also known as celestial plane. Note that the 
coordinates $\alpha$ and $\beta$ corresponds to the apparent perpendicular distance of the image as 
seen from the axis of symmetry and the apparent perpendicular distance of the image from its 
projection on to the equatorial plane, respectively. With this construction, we obtain the projection 
of wormhole's throat on the observer's sky. The celestial coordinates for a distant observer can be
written \cite{Bardeen:1973gb} in the form 
\begin{eqnarray}
\label{alp-bet}
\alpha &=& \lim_{r \rightarrow \infty} \left(-r^2 \sin \theta_{0} \frac{d \phi}{dr} \right),
\nonumber \\
\beta &=& \lim_{r \rightarrow \infty} \left(r^2 \frac{d \theta}{dr} \right),
\end{eqnarray}
where $\theta_{0}$ represents the inclination angle between the rotation axis of the wormhole and the 
direction to the observer. By using Eqs.~(\ref{u}), (\ref{u1}), (\ref{quant}), and (\ref{alp-bet}), the 
celestial coordinates transform into 
\begin{eqnarray}
\label{alp-bet1}
\alpha &=& -\xi \csc \theta_{0}, \nonumber \\
\beta &=& \pm \sqrt{\eta +a^2 \cos^2 \theta_{0} -\xi^2 \cot^2 \theta_{0}}.
\end{eqnarray}
It is clear from Eq.~(\ref{alp-bet1}) that the shapes of the Kerr-like wormhole's shadow depend not 
only on its spin $a$ but also on its inclination angle $\theta_0$. When the observer is located in the 
equatorial plane ($\theta_{0}=\pi /2$) of the wormhole,  the expressions in Eq.~(\ref{alp-bet1}) reduce 
to
\begin{eqnarray}\label{eqtc}
\alpha &=& -\xi, \nonumber\\ 
\beta &=& \pm \sqrt{\eta}.
\end{eqnarray}
\begin{figure}
  \includegraphics[scale=0.65]{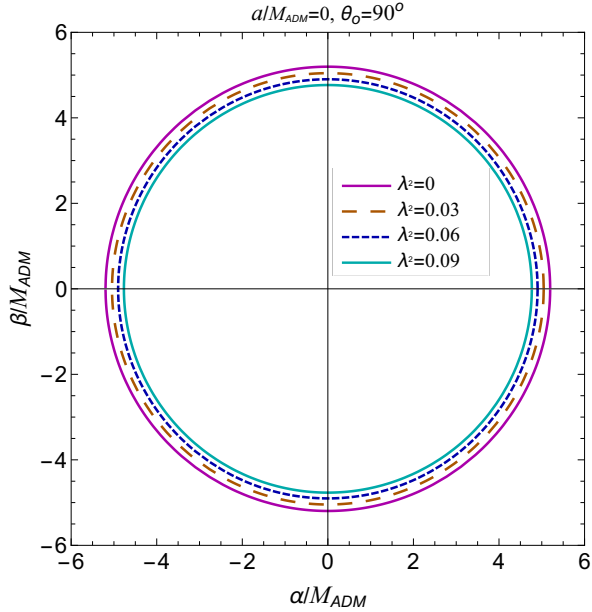}
  \caption{\label{nshad} Plot illustrating the shadow images of the nonrotating wormholes by varying 
  the deviation parameter $\lambda^2$. Here $\lambda^2=0$ corresponds to the Schwarzschild black hole 
  case ($M_{ADM}$ is set equal to $1$ for both of the spacetimes).}
\end{figure}
A dark region in the observer's sky can be obtained when the impact parameters $\eta$ and $\xi$ 
attains all the possible values. This dark region is known as the shadow of the wormhole. The 
nonrotating wormhole shadow can be constructed by the following expression
\begin{eqnarray}\label{nonr}
\alpha^2 + \beta^2 = \frac{2 r^2 \left(r^2-3 M^2\right)}{(M-r)^2},
\end{eqnarray}
which implies that the shadow casts by a nonrotating wormhole ($a=0$) has a circular shape with a 
radius $\sqrt{\frac{2 r^2 \left(r^2-3 M^2\right)}{(M-r)^2}}$. In order to plot the shadows, we consider 
the ADM mass ($M_{ADM}$) of the wormhole. The shadows of nonrotating wormhole can be seen from 
Fig.~\ref{nshad}. The figure shows that the radius of shadow decreases with a small change in deviation 
parameter $\lambda^2$. Moreover, the radius of shadow is smaller than that of Schwarzschild black hole 
(cf. Fig.~\ref{nshad}). In contrast for a rotating wormhole the sum of $\alpha^2 + \beta^2$, takes the 
complicated form
\begin{eqnarray}\label{rot}
\alpha^2 + \beta^2 &=& \frac{a^2 (M + r)^2 +2 r^2 (r^2 -3 M^2)}{(M-r)^2}.
\end{eqnarray}
The presence of spin $a$ in the expression (\ref{rot}) results in manifestly different shapes of the 
shadow of the Kerr-like wormhole in comparison with the nonrotating one. It is noticeable that the 
expression (\ref{rot}) depends on deviation parameter $\lambda^2$ while the ADM mass of the wormhole is 
considered. This is because as we pointed out we shall refer to the ADM mass as the physical mass of the 
wormhole viewed by an observer located at spatial infinity. In order to see it more clearly, we construct 
shapes of the Kerr-like wormholes shadows for nonzero values of the spin $a$, the deviation parameter 
$\lambda^2$, and the inclination angle $\theta_0$. We plot the shadow images of the Kerr-like wormholes 
for different values of the spin $a$, the deviation parameter $\lambda^2$, and the inclination angle 
$\theta_0$ (cf. Figs.~\ref{shad} and \ref{shad-1}). Interestingly, we find that the shape of the shadow 
is oblate instead of being circular when compare with the nonrotating case. This oblateness or distortion 
in the shape arises because of the nonzero values of the spin $a$. In Fig.~\ref{shad}, we show the 
variation of the spin $a$ while the Fig.~\ref{shad-1} depicts the effect of deviation parameter 
$\lambda^2$ on the shape of the shadow. It is clearly observed that the presence of the deviation 
parameter $\lambda^2$ influences the radius of the shadow.

An increase in the magnitude of spin $a$ increases the amount of distortion in the shadow of the 
Kerr-like wormhole (cf. Figs.~\ref{shad} and \ref{shad-1}). On the other hand when we fix the values of 
$a$ and at the same stage we do a change in $\lambda^2$ values that also increases the distortion in the 
shape. The shape is maximally distorted when the observer is located in equatorial plane as can be 
seen from Fig.~\ref{shad-1}. While we talk about the effect on the size or radius of the Kerr-like 
wormhole's shadow, we find that the radius is continuously decreasing with a small change in the 
magnitude of the deviation parameter $\lambda^2$ (cf. Fig.~\ref{shad-1}). We also include the case of 
the Kerr black hole ($\lambda^2 =0$) by considering the same ADM mass ($M_{ADM} =1$) to see the 
difference in their shadow images (cf. Fig.~\ref{shad-1}). We discover that the shadow images of the 
Kerr-like wormholes are smaller in comparison to that of the Kerr black hole. In case of the Kerr 
spacetime, one can only observe the effect of spin $a$ on shadow, but in our case we have an additional 
parameter $\lambda^2$ which provides more images of the shadow so that extract more substantial 
information from them. Note that the actual size of the wormhole's shadow is bigger as compare to 
wormhole's geometrical size. Since the wormhole bends photon trajectories, therefore, the actual cross 
section of the shadow is greater than that of the geometrical one.

\section{Conclusion}
\label{conclusion}
The discovery of quasar in 1960s  provided compelling evidence for the existence of supermassive black 
holes located at least in some galactic binary systems in the centers of most of the large galaxies, 
e.g., the Milky Way and Messier 87 ostensibly  have such objects, namely, Sgr A$^*$ and M87$^*$. 
Supermassive black holes are considered as astrophysical black holes and they form from the 
gravitational collapse of matter. Besides, the detection of ringdown frequencies from the black holes 
provide precise tests that astrophysical black holes are accurately described by the Kerr spacetime. 
Motivated by these tests, we construct the theoretical analysis of the shadow for the Kerr-like 
wormholes. Therefore, observational investigation of the Kerr-like wormhole shadows will be very 
interesting and a useful tool to demonstrate the true nature of it. The observation of the shadow also 
provides a tentative way to determine the parameters governing the evolution of  the wormholes. As it 
is well known one of the approaches to detect supermassive black holes is based on high resolution 
imaging of them. The Event Horizon Telescope which functions on the Very Long Baseline Interferometry 
technique, has been designed to resolve the problem of the detection of the supermassive black hole at 
the center of the Milky Way. This technique is able to achieve the angular resolution comparable to 
sub-millimeter wavelength diapason. 

In this paper, we concentrate on the construction of the shadow cast by the Kerr-like wormholes. We 
have evaluated the test particle geodesics and determined the trajectories of photons around the 
Kerr-like wormhole. It has been noticed that the photons approaching the wormhole with sufficient 
angular momentum form unstable spherical orbits around it. In further investigation, we introduced the 
impact parameters that define the boundary of the shadow against the bright background. In order to 
visualize the optical images of the shadow of the Kerr-like wormhole, the celestial coordinates are 
evaluated. We have plotted these coordinates into the celestial plane to get the shadow images, 
eventually discussed the nonrotating and rotating cases of the wormhole's shadows. We have discovered 
that the radii of the nonrotating wormhole's shadows are smaller in comparison to that of the 
Schawrzschild black hole. We have obtained different shapes of the shadow for the Kerr-like wormholes 
by varying the parameters. We find  that the shapes of the Kerr-like wormhole's shadow are oblate or 
distorted. Mainly, this distortion occurs due to the magnitude of the spin parameter $a$ and inclination 
angle $\theta_0$. The shapes are more distorted for the extreme value of the spin $a$ and when the 
observer is located into the equatorial plane. In addition, it is found that the radius of the shadow 
images is also sensitive to value of the deviation parameter $\lambda^2$; a small change in $\lambda^2$ 
decreases the radius of the shadow substantially. As a consequence the radius decreases with deviation 
parameter $\lambda^2$ in general. Our results provide the deviations when a comparison with the Kerr 
spacetime is taking on account by using the same ADM mass for both of the spacetimes. We have also 
discovered that the radii of the Kerr-like wormhole's shadows are smaller in comparison to that of the 
Kerr black hole. This important study will be helpful to extract the essential information about the 
Kerr-like wormhole and its existence as the astrophysical object. We are expecting that in upcoming years 
the direct images of wormholes will be observe by the Event Horizon Telescope. 

\acknowledgments
M.A., A.B., and S.H. would like to thank University of KwaZulu-Natal and the National Research 
Foundation for financial support. We would like to thank the referees for useful comments and 
suggestions.

\end{document}